\documentclass[review, 11pt,sort&compress]{elsarticle}
\usepackage[utf8]{inputenc}
\usepackage[toc,page]{appendix}
\usepackage{amssymb}            
\usepackage{babel}
\usepackage{graphicx}
\usepackage{booktabs}
\usepackage{xcolor}
\usepackage{gensymb} 
\usepackage{amsmath,amsfonts,amsthm,bm,mathtools} 
\usepackage{mathrsfs} 
\usepackage{geometry}
\usepackage{gensymb}
\usepackage{multirow}
\usepackage[skip=6pt plus1pt, indent=40pt]{parskip}
\usepackage{setspace} 

\journal{Corrosion Science}
\usepackage{graphicx}
\graphicspath{{Figures/}}
\usepackage{caption}
\usepackage{subcaption}
\usepackage{float}
\usepackage{hyperref}
\hypersetup{colorlinks=true, linkcolor=blue, citecolor=black, filecolor=magenta, urlcolor=black}
\usepackage{soul}
\usepackage{accents}
\usepackage{color,soul} 
\usepackage{color} 
\usepackage{xcolor,soul}
\sethlcolor{lightgray}

\usepackage{upgreek}

\begin{document}
\begin{frontmatter}
\title{A microstructure-sensitive electro-chemo-mechanical phase-field model of pitting and stress corrosion cracking}

\author[1]{Maciej Makuch}
\author[2]{Sasa Kovacevic}
\author[1]{Mark R. Wenman}
\author[2]{Emilio Mart\'inez-Pa\~neda \corref{cor1}}
\ead{emilio.martinez-paneda@eng.ox.ac.uk}

\cortext[cor1]{Corresponding authors}

\address[1]{Imperial College London, Centre for Nuclear Engineering, South Kensington Campus, London SW7 2AZ, UK}
\address[2]{Department of Engineering Science, University of Oxford, Oxford OX1 3PJ, UK}

\begin{abstract}

\noindent An electro-chemo-mechanical phase-field formulation is developed to simulate pitting and stress corrosion in polycrystalline materials. The formulation incorporates dependencies of mechanical properties and corrosion potential on crystallographic orientation. The model considers the formation and charging dynamics of an electric double layer through a new general boundary condition for the solution potential. The potential of the model is demonstrated by simulating corrosion in polycrystalline materials with various grain morphology distributions. The results show that incorporating the underlying microstructure yields more extensive defects, faster defect kinetics, and irregular pit and crack shapes relative to a microstructurally-insensitive homogeneous material scenario.\\

\end{abstract}

\end{frontmatter}
\cleardoublepage

\section{Introduction} \label{sec1}

Corrosion is a common cause of failure of engineering components and structures across various industries \cite{Thompson2007, Liu2023}. While physical and electrochemical measures can prevent uniform corrosion, pitting corrosion is more damaging and difficult to detect and predict \cite{Uhlig2008}. This type of corrosion often occurs after the local breakdown of the passive film on the metal surface and governs the design and service life of engineering components operating in aggressive environments. In the presence of mechanical loading, formed pits can create hotspots for stress concentration, resulting in cracks and premature failure, which can lead to catastrophic events \cite{Marcus2002}.

The significance of metallurgy and how the interplay between metallurgy and the corrosive environment influence pitting corrosion is still not completely understood \cite{Frankel1998, Eliyan2014}. Pitting initiates at a microscopic level and is influenced by harsh environments and microstructural features of the corroding material. Experimental studies have shown that the onset and propagation rate of corrosion pits strongly depend on the crystallographic orientation of the material \cite{Sato1996, Wang2014, FU2019, SCHREIBER2006}. For instance, it has been found that planes with \{111\} orientations exhibit the highest resistance to localized corrosion in stainless steel \cite{Lindell2015, SHAHRYARI2009}. At the same time, the opposite tendency has been reported for aluminum alloys \cite{Yasuda1990, Koroleva2007}. Moreover, in the presence of mechanical loading, the local mechanical properties affect the crack initiation and crack propagation \cite{QIAO2011}. The synergistic effect of mechanical loading, the underlying microstructure of the corroding material, and a corrosive environment can significantly reduce the corrosion resistance and mechanical integrity of materials \cite{MOORE2023,cui2022generalised}.

Observing pitting events and quantifying their kinetics in experiments is challenging due to the intrinsic localized behavior and length scale of the problem. Numerical models that resolve the underlying physical and chemical processes can be utilized to predict the corrosion behavior of materials and to guide their design. Tracking the evolution of the metal-environment interface under mechano-electrochemical effects is a complex computational challenge due to the strongly coupled nature of the problem. A wide variety of numerical techniques have been proposed to track the evolution of the corrosion front at different scales, including arbitrary Lagrangian-Eulerian approaches \cite{SARKAR2012, SUN2014, Brewick2017, Brewick2019}, level set methods \cite{Duddu2014, Vagbharathi2014}, and peridynamics \cite{Jafarzadeh2019, CHEN2015}. While these computational tools have merit and consider microstructural features \cite{Brewick2017, Brewick2019}, they are limited in coupling with other physicomechanical phenomena and handling geometric interactions in arbitrary dimensions (2D/3D), such as the coalescence of corrosion pits and cracks.

Phase-field formulations have emerged as a promising approach for the computational treatment of topological changes and quantitative modeling of interfacial phenomena at different length scales \cite{Chen2002}. In phase-field models, the interface between two phases (e.g., environment and metal) is smoothed over a thin diffuse region using a continuous auxiliary field variable (e.g., $\phi$), see Fig. {\ref{Fig1}}. The phase-field variable $\phi$ has a distinct value in each phase (e.g., $\phi = 0$ in the environment and $\phi = 1$ in the metal) while varying smoothly in between. The motion of the interface is implicitly tracked without presumptions or prescribing the interface velocity. Such interface representation enables capturing topological changes of arbitrary complexity (e.g., divisions or merging of interfaces) in arbitrary domains without requiring any special treatments or \textit{ad hoc} criteria. Recent developments in phase-field modeling of corrosion damage have shown that the phase-field method can naturally track the evolution of the metal-environment interface and capture pit-to-crack transition and crack propagation in arbitrary domains \cite{MAI2016, Mai2018, Tsuyuki2018, Ansari2018, Ansari2019, LIN2019, LIN2021, CUI2021, CUI2023, KOVACEVIC2023}. Recently, the phase-field method has been extended to include the role of crystallographic orientation in single crystals \cite{Sahu2022, SAHU2023} and polycrystalline materials \cite{Chadwick2018, Chen2022, Brewick2022, Nguyen2017}. However, the electrochemistry of the system and the synergistic effect of aggressive environments, underlying microstructures, and mechanical loading are not fully resolved in these models \cite{Sahu2022, SAHU2023, Chadwick2018, Chen2022, Brewick2022, Nguyen2017}. Moreover, electrochemical interfacial phenomena associated with the formation of an electric double layer (EDL) at the metal-electrolyte interface and its role in the transport of ionic species and solution potential distribution \cite{Schmickler2020} have not been considered so far in corrosion models despite their importance \cite{SARKAR2012, SUN2014, Brewick2017, Brewick2019, Duddu2014, Vagbharathi2014, Jafarzadeh2019, CHEN2015, MAI2016, Mai2018, Tsuyuki2018, Ansari2018, Ansari2019, LIN2019, LIN2021, CUI2021, CUI2023, KOVACEVIC2023, Sahu2022, SAHU2023, Chadwick2018, Chen2022, Brewick2022, Nguyen2017}. In existing phase-field models, the solution potential distribution is typically related to the interface propagation rate while the formation and charging dynamics of the EDL are neglected \mbox{\cite{Chadwick2018, Ansari2018, LIN2019, LIN2021, Chen2022, Tsuyuki2018, CUI2023, Brewick2022}}. This work constitutes the first electro-chemo-mechanical phase-field formulation for assessing pitting and stress-assisted corrosion in polycrystalline materials. The model captures the sensitivity to the crystallographic orientation and inherent electrochemical phenomena occurring at the metal-electrolyte interface. A general boundary condition for the solution potential that accounts for the formation and charging dynamics of the EDL is proposed and implemented into phase-field corrosion models. The condition is derived from a resistor-capacitor equivalent circuit model and allows the inclusion of the effects of the EDL without explicitly simulating it in the computational domain. 

The remainder of the paper is organized as follows. The underlying electrochemistry and corrosion mechanisms that govern pitting and stress-assisted corrosion in polycrystalline materials are presented in the following section. The diffuse interface (phase-field) model is subsequently developed based on a generalized thermodynamic free energy functional that incorporates chemical, gradient, electric, and mechanical contributions. Microstructural influence is integrated into the model through dependencies of mechanical properties and corrosion potential on crystallographic orientation. The governing equation for the solution potential is supplemented with a boundary condition that accounts for the formation and charging dynamics of an EDL.  The constructed formulation is calibrated and validated against experimental measurements on a thin stainless steel wire immersed in a corrosive environment in Section \ref{sec3}. Two case studies are presented in Section \ref{sec4}, demonstrating the ability of the model to capture pitting and stress-assisted corrosion in polycrystalline materials: pitting corrosion initiated after the local breakdown of a protective layer and a single edge notch specimen subjected to tension. Conclusions of the study, along with recommendations for future work, are summarized in Section \ref{sec5}.

\section{Phase-field formulation} \label{sec2}
\subsection{Underlying electrochemistry} \label{sec21}

Pitting corrosion initiates after the local breakdown of the protective film, exposing the fresh metallic surface to a corrosive environment. Neglecting the precipitation of salts and stable phases at the exposed metal surface, the corrosion process can be summarized with the following reactions \cite{Uhlig2008}
\begin{equation} \label{eqn1}
    \text{M}_{(s)} \rightarrow \text{M}_{(aq)}^{z_1+} + z_1e^- \text{ (metal dissolution)}
\end{equation}
\begin{equation} \label{eqn2}
    \text{M}_{(aq)}^{z_1+} + \text{H}_2\text{O} \xrightleftharpoons[k_{1b}]{k_{1f}} \text{M}(\text{OH})^{(z_1-1)+} + \text{H}^+ \text{ (primary hydrolysis)}
\end{equation}
\begin{equation} \label{eqn3}
    \text{H}_2\text{O} \xrightleftharpoons[k_{2b}]{k_{2f}} \text{OH}^- + \text{H}^+ \text{ (water dissociation)}
\end{equation}
where M is the corroded metal, $z_1$ is the charge number, and $k_{1f}$, $k_{2f}$, $k_{1b}$, and $k_{2b}$ are the corresponding rate constants for the forward and backward reactions respectively. 

\begin{figure}[h!]
    \centering
    \includegraphics[width = 16 cm]{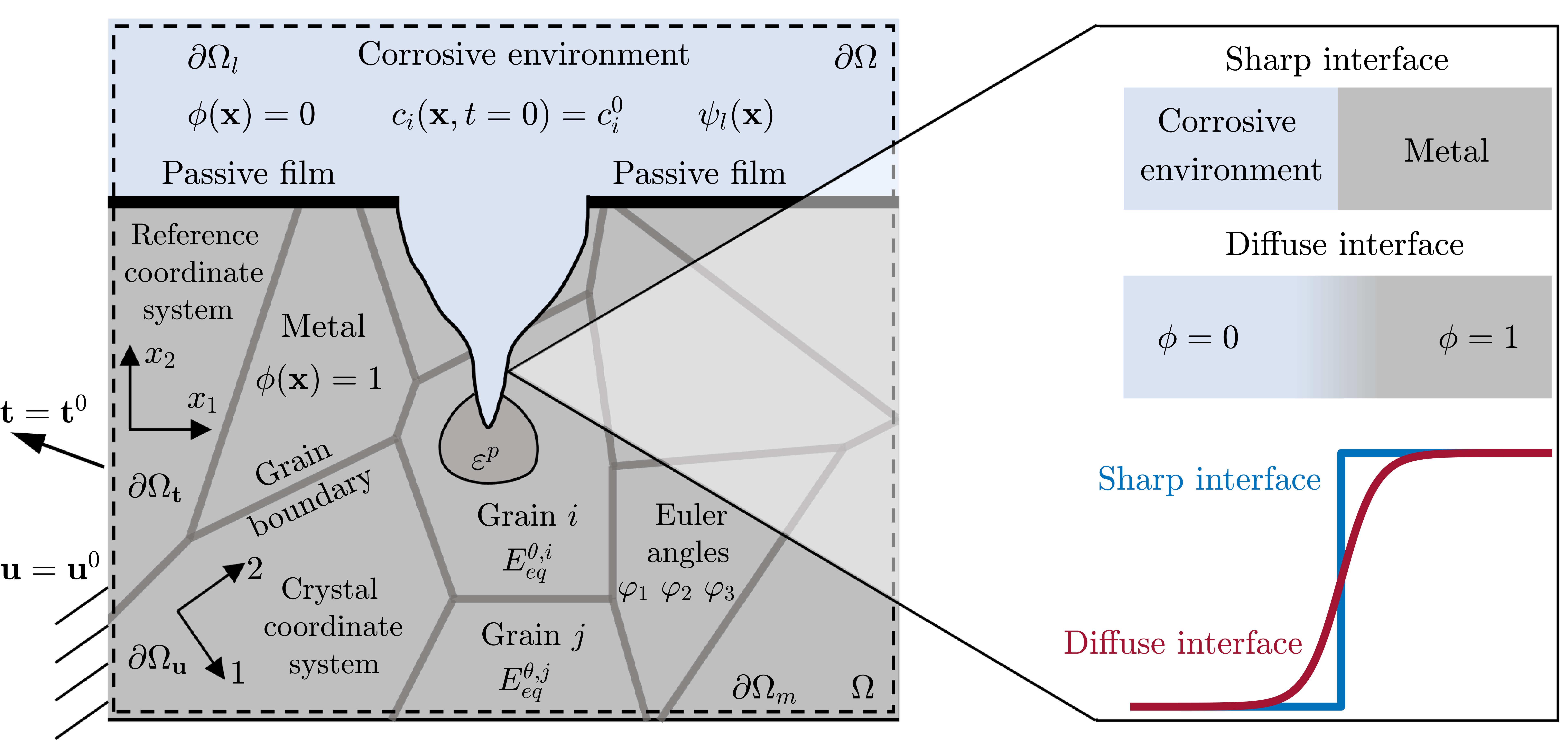}
    \captionsetup{labelfont = bf,justification = raggedright}
    \caption{Polycrystalline material in contact with corrosive environment and diffuse interface description of the liquid (electrolyte $\phi$ = 0) and solid (electrode $\phi = 1$) phases.}
    \label{Fig1}
\end{figure}

The electrochemical system considered includes an electrode and an ionic conducting electrolyte, as illustrated in Fig. \ref{Fig1}. It is assumed that the aqueous electrolyte is a NaCl-based solution. The above reactions and the surrounding environment are described using a set of concentrations of ionic species represented by $\overrightarrow{c} = (c_1 = \text{M}^{z_1+}, c_2 = \text{M(OH)}^{(z_1-1)+}, c_3 = \text{H}^+, c_4 = \text{OH}^-, c_5 = \text{Na}^+, c_6 = \text{Cl}^-)$. While Na$^+$ and Cl$^-$ ions do not participate in the reactions in Eqs. ({\ref{eqn1}}), ({\ref{eqn2}}), and ({\ref{eqn3}}), they influence the movement and distribution of ions in solution since they are charged ions. The electrode is a polycrystalline material composed of grains with different orientations separated by grain boundaries (GBs), as shown in Fig. \ref{Fig1}. The orientation of each grain with respect to the reference coordinate system is described using the Euler angles $\varphi_1$, $\varphi_2$, and $\varphi_3$, see Fig. \ref{Fig1}. It is emphasized that the corrosion potential and mechanical properties of each grain are dependent on its crystallographic orientation. The GBs in the present model do not have a physical character, an approach frequently followed in the literature \cite{Brewick2017, Brewick2019}. They are included to differentiate the corrosion potential and mechanical properties between grains. The role of GBs in pitting corrosion will be addressed in future work.

A continuous phase-field variable $\phi$ is introduced to distinguish between the metal and the liquid phase: $\phi = 1$ represents the solid phase, $\phi = 0$ corresponds to the liquid phase, and $0 < \phi < 1$ indicates the thin interfacial region between the phases (solid-liquid interface). It is assumed that no diffusion occurs across the domain boundary such that the normal fluxes vanish on the boundary ($\mathbf{n} \cdot \mathbf{J} = 0$). The independent kinematic variables include the phase-field parameter that describes the evolution of the corroding interface $\phi(\mathbf{x},t)$, the displacement vector $\mathbf{u}(\mathbf{x},t)$ that characterizes the deformation of the material, the concentration variable $c_{i}(\mathbf{x},t)$ for each ionic species considered, and the solution potential $\psi_l(\mathbf{x},t)$.

\subsection{Thermodynamics} \label{sec22}

The free energy functional of a heterogeneous system (Fig. \ref{Fig1}) is defined as
\begin{equation} \label{eqn4}
                   \mathscr{F} = \int_\Omega \Big[f^{chem}(\overrightarrow{c},\phi) + f^{grad}(\nabla\phi) + f^{elec}(\overrightarrow{c},\psi_l) + f^{mech}(\nabla \mathbf{u},\phi)\Big]\,d\Omega
\end{equation}
and includes the contribution from the chemical $f^{chem}$, gradient $f^{grad}$, electric $f^{elec}$, and mechanical $f^{mech}$ free energy densities that are defined below.

\subsubsection{Chemical free energy density} \label{sec221}

The chemical free energy density of the metal and the electrolyte is represented as a two-phase mixture 
\begin{equation} \label{eqn5}
                   f^{chem}(\overrightarrow{c},\phi) = f^{chem}_1(\bar{c}_1, \phi) + \sum_{i = 2} f^{chem}_{i}(\bar{c}_i) + \omega g(\phi)
\end{equation}
where $f^{chem}_1(\bar{c}_1, \phi)$ and $f^{chem}_{i}(\bar{c}_i)$ ($i\neq1$) are the chemical free energy densities as a function of normalized ion concentrations $\bar{c}_i = c_i V_m$, where $V_m$ stands for the molar volume of the metal. In Eq. (\ref{eqn5}), $g(\phi) = 16\phi^2(1-\phi)^2$ is the double-well free energy function employed to describe the two equilibrium states for the electrode and electrolyte, and $\omega$ is the constant that determines the energy barrier at $\phi = 1/2$ between the two minima at $\phi = 0$ and $\phi = 1$. The chemical free energy density associated with the metal ion concentration $f^{chem}_1(\bar{c}_1,\phi)$ is defined as the sum of the free energy densities of the liquid and solid phases \cite{KKS1999}
\begin{equation} \label{eqn6}
                   f^{chem}_1(\bar{c}_1,\phi) = (1-h(\phi))f^{chem}_{l}(\bar{c}_1^l) + h(\phi)f^{chem}_{s}(\bar{c}_1^s)
\end{equation}
where $f^{chem}_{l}(\bar{c}_1^l)$ and $f^{chem}_{s}(\bar{c}_1^s)$ are the chemical free energy densities as a function of normalized metal ion phase-concentrations within the liquid $\bar{c}_1^l$ and the solid phases $\bar{c}_1^s$. The function $h(\phi) = \phi^3(10-15\phi+6\phi^2)$ is a monotonically increasing interpolation function. The interfacial region is assumed to be a mixture of solid and liquid with different concentrations but with the same diffusion chemical potential \cite{KKS1999}
\begin{equation} \label{eqn7}
                   \bar{c}_1 = (1-h(\phi))\bar{c}_1^l + h(\phi)\bar{c}_1^s \quad\mathrm{}\quad \frac{\partial f^{chem}_{l}(\bar{c}_1^l)}{\partial \bar{c}_1^l} = \frac{\partial f^{chem}_{s}(\bar{c}_1^s)}{\partial \bar{c}_1^s}
\end{equation}
The chemical free energy densities within the pure phases are commonly approximated by parabolic functions around equilibrium concentrations with the same density curvature parameter $A$ as \cite{MAI2016}
\begin{equation} \label{eqn8}
                   f^{chem}_{l}(\bar{c}_1^l) = \frac{1}{2}A(\bar{c}_1^l - \bar{c}_1^{l,eq})^2  \quad\mathrm{}\quad  f^{chem}_{s}(\bar{c}_1^s) = \frac{1}{2}A(\bar{c}_1^s - \bar{c}_1^{s,eq})^2
\end{equation}
where $\bar{c}_1^{l,eq} = c_1^{l,eq} V_m$ and $\bar{c}_1^{s,eq} = c_1^{s,eq} V_m = 1$ are the normalized equilibrium phase concentrations in the liquid and solid phases. Alternatively, the chemical free energy density can be approximated assuming a dilute solution \cite{LIN2021} or determined using the first-principles calculations and thermodynamic databases \cite{KUMARTHAKUR2023}. The equilibrium phase-concentration in the liquid phase $c_1^{l,eq}$ is determined based on the solubility of salts formed on the exposed metal surface as the formation of corrosion products and the passive film are neglected in the present corrosion mechanism, Eqs. (\ref{eqn1}), (\ref{eqn2}), and (\ref{eqn3}). Combining Eqs. (\ref{eqn7}) and (\ref{eqn8}) returns the following expression for the chemical free energy density associated with the metal ion concentration
\begin{equation} \label{eqn9}
                   f^{chem}_1(\bar{c}_1,\phi) = \frac{1}{2} A \Big[\bar{c}_1 -h(\phi) (\bar{c}_1^{s,eq} - \bar{c}_1^{l,eq}) - \bar{c}_1^{l,eq}\Big]^2
\end{equation}
The contribution to the chemical free energy density from the other ions present in solution ($i\neq1$) is expressed following the dilute solution theory \cite{Bazant2013}
\begin{equation} \label{eqn10}
                   f^{chem}(\bar{c}_i) = \frac{RT}{V_m}\sum_{i = 2} \bar{c}_i\ln \bar{c}_i + \sum_{i = 2} \bar{c}_i \mu_{i}^{\Theta}
\end{equation}
where $\mu_{i}^{\Theta}$ is the reference chemical potential of ionic species $i$, $R$ is the universal gas constant, and $T$ is the absolute temperature.

\subsubsection{Interfacial free energy density} \label{sec222}

The interfacial free energy density is defined as
\begin{equation} \label{eqn11}
                   f^{grad}(\nabla\phi) =\frac{1}{2}\kappa|\nabla \phi|^2
\end{equation}
where $\kappa$ is the isotropic gradient energy coefficient. The phase-field parameters $\omega$ (Eq. (5)) and $\kappa$ are connected to the interfacial energy $\Gamma$ and the interface thickness $\ell$ as \cite{KOVACEVIC2020}
\begin{equation} \label{eqn12}
                   \omega =\frac{3\Gamma}{4\ell}  \quad\mathrm{}\quad \kappa = \frac{3}{2}\Gamma\ell
\end{equation}
\subsubsection{Electric free energy density} \label{sec223}

The electric free energy density of the system subjected to a net electric potential in the electrolyte can be written as \cite{Jefimenko1989}
\begin{equation}\label{eqn13}
    f^{elec}(\overrightarrow{c},\psi_l) =  \rho \psi_l \quad\mathrm{}\quad  \rho = F\sum_{i}z_i c_i
\end{equation}
where $F$ is Faraday's constant, $\rho$ is the local charge density, and $z_i$ is the charge number of component $i$. 

\subsubsection{Mechanical free energy density} \label{sec224}

The total mechanical free energy density is additively decomposed into elastic $f_e^{mech}$ and plastic $f_p^{mech}$ components
\begin{equation} \label{eqn14}
                   f^{mech}(\nabla \mathbf{u},\phi) = h(\phi) (f_e^{mech} + f_p^{mech})
\end{equation}
where $h(\phi)$, defined in Section {\ref{sec221}}, acts as the degradation function to ensure the transition from the intact solid (uncorroded material) to the completely corroded (liquid) phase. The elastic strain energy density $f_e^{mech}$ is a quadratic form of the elastic strain
\begin{equation} \label{eqn15}
                   f^{mech}_e(\nabla \mathbf{u}, \phi) = \frac{1}{2} \bm{\varepsilon}^{e} : \mathbf{C} : \bm{\varepsilon}^{e} \quad\mathrm{}\quad \bm{\varepsilon}^{e}=\bm{\varepsilon} - \bm{\varepsilon}^{p}
\end{equation}
where $\mathbf{C}$ is the rank-four elastic stiffness tensor. The elastic strain tensor $\bm{\varepsilon}^{e}$ is obtained by subtracting the plastic strain tensor $\bm{\varepsilon}^{p}$ from the total strain tensor $\bm{\varepsilon}$. The total strain tensor is the symmetric part of the displacement gradient
\begin{equation} \label{eqn16}
                   \bm{\varepsilon} = \frac{1}{2} (\nabla \mathbf{u} + (\nabla \mathbf{u})^T)
\end{equation}
It is assumed that the material exhibits cubic symmetry so that the rank-four elastic stiffness tensor, given in the principal material system, reads
\begin{equation} \label{eqn17}
                   C_{ijkl} = C_{12}\delta_{ij}\delta_{kl}+C_{44}(\delta_{ik}\delta_{jl}+\delta_{il}\delta_{jk})+(C_{11}-C_{12}-2C_{44})\delta_{ijkl}
\end{equation}
where $C_{11}$, $C_{12}$, and $C_{44}$ are the three monocrystal elastic constants. The elastic stiffness tensor is transformed into the global coordinate system using the standard tensor transformation
\begin{equation} \label{eqn18}
                   \mathbf{C}^{\prime} = \mathbf{R} \cdot \mathbf{R} \cdot \mathbf{C} \cdot \mathbf{R}^T \cdot \mathbf{R}^T
\end{equation}
where $\mathbf{R}$ is the transformation matrix defined based on the Euler angles $\varphi_1$, $\varphi_2$, and $\varphi_3$ of each grain.

The plastic strain energy density $f_p^{mech}$ is incrementally computed from the plastic strain tensor $\bm{\varepsilon}^{p}$ and the Cauchy stress tensor $\bm{\sigma_0}$ for the intact configuration
\begin{equation} \label{eqn19}
                   f_p^{mech} = \int_0 ^t \bm{\sigma_0} : \dot{\bm{\varepsilon}}^{p}\,d t
\end{equation}
where $\dot{\bm{\varepsilon}}^{p}$ is the effective plastic strain rate. The constitutive behavior of the material is characterized by following von Mises plasticity theory \cite{Simo1998}. The material work hardening is defined by assuming an isotropic power-law hardening behavior relating the flow stress $\mathbf{\sigma}$ and equivalent plastic strain $\mathbf{\varepsilon}^{p}$ as
\begin{equation} \label{eqn20}
                   \sigma = \sigma_y \Big(1 + \frac{\mathbf{\varepsilon}^p}{\varepsilon_0}\Big)^N \quad\mathrm{}\quad \mathbf{\varepsilon}^{p} = \sqrt{\frac{2}{3}\bm{\varepsilon}^p : \bm{\varepsilon}^p}
\end{equation}
where $\sigma_y$ is the yield stress, $\varepsilon_0$ is the yield strain, and $N$ is the strain hardening exponent ($0\le N \le 1$). It should be noted that, although the mechanical behavior of the metal is characterized following the von Mises plasticity theory for simplicity, the proposed framework can readily be extended to incorporate crystal plasticity \cite{Song2023,lucarini2024umat4comsol}.

\subsection{Governing equations} \label{sec23}

The evolution of the phase-field parameter is assumed to obey the Allen-Cahn equation for non-conserved fields \cite{ALLEN19791085}
\begin{equation} \label{eqn21}
\frac{\partial \phi}{\partial t} = - L\frac{\delta \mathscr{F}}{\delta \phi} = -L\Big(\frac{\partial f^{chem}}{\partial\phi} - \kappa\nabla^2\phi \Big)\quad \text{in}\quad \Omega \quad\mathrm{} \kappa \mathbf{n} \cdot \nabla \phi = 0 \quad \text{on}\quad\partial\Omega
\end{equation}
where $L > 0$ is the kinetic coefficient that characterizes the interfacial mobility. The mechanical term in the above equation, $\partial f^{mech}/\partial \phi = h^{\prime} (\phi)f^{mech}$, is neglected as it would introduce nonphysical couplings under diffusion-controlled corrosion conditions \cite{CUI2021, CUI2023, KOVACEVIC2023}. In the present model, the role of mechanical fields on the interface kinetics is incorporated \textit{via} a mechano-electro-chemical coupling as described below in Section \ref{sec24}.

The transport of ionic species in the system is subjected to mass balance consideration
\begin{equation} \label{eqn22}
\frac{\partial c_i}{\partial t} = - \nabla \cdot \mathbf{J}_i + R_i\quad \text{in}\quad \Omega \quad\mathrm{} \mathbf{n}\cdot \mathbf{J}_i = 0 \quad \text{on}\quad\partial\Omega
\end{equation}
where $\mathbf{J}_i$ stands for the electrochemical flux and $R_i$ denotes the volumetric chemical reaction rates associated with the primary hydrolysis and water dissociation reactions given in Eqs. ({\ref{eqn2}}) and ({\ref{eqn3})}. The simplest linear constitutive law yields
\begin{equation} \label{eqn23}
\mathbf{J}_i = - M_i\nabla \Big(\frac{\delta \mathscr{F}}{\delta c_i}\Big) \quad\mathrm{}\quad M_i > 0
\end{equation}
Expressing the mobility parameter $M_i$ with the Nernst–Einstein equation ($M_i = D_i/(\partial^2f^{chem}/\partial \bar{c}_i^2)$) renders an extended diffusion equation for the metal ion and the standard Nernst–Planck equation for the other ionic species in the system
\begin{equation} \label{eqn24}
\left\{
\begin{aligned}
& \frac{\partial c_1 }{\partial t} = -\nabla \cdot \mathbf{J}_1 + R_1 \quad\mathrm{}\quad \mathbf{J}_1 = -D_1\Big(\nabla c_1 - h^{\prime}(\phi)(c_1^{s,eq} - c_1^{l,eq}) \nabla \phi + \frac{z_1 F}{R T} c_1 \nabla \psi_l \Big)\\
& \frac{\partial c_i}{\partial t} = -\nabla \cdot \mathbf{J}_i + R_i \quad\mathrm{}\quad \mathbf{J}_i = -D_i \Big(\nabla c_i + \frac{z_i F}{R T} c_i \nabla \psi_l \Big) \quad\mathrm{}\quad i \neq 1\\
\end{aligned}
\right\}\
\end{equation}
Following Refs. \cite{Ansari2018, Mai2018, CUI2023}, the electromigration flux in the equation for the metal ion is derived assuming the dilute solution theory. $D_i$ is the effective diffusion coefficient interpolated with the phase-field parameter between the phases: $D_{i} = D^s_{i}h(\phi)+(1-h(\phi))D^l_{i}$, where $D^l_{i}$ and $D^s_{i}$ are the diffusion coefficients of ions in the liquid and solid phases. Here, $D^s_{i} \ll D^l_{l}$ is enforced to prevent the diffusion of ions within the solid phase. Assuming that the chemical equilibrium is instantaneously satisfied in the electrolyte, the volumetric chemical reaction rates $R_i$ are
\begin{equation} \label{eqn25}
\begin{aligned}
R_1 = k_{1b}(c_2 c_3-K_1 c_1) \quad\mathrm{}\quad R_2 = -R_1 \quad\mathrm{}\quad R_3 = R_2 + R_4 \quad\mathrm{}\quad R_4 = k_{2b}(K_2 - c_3 c_4)
\end{aligned}
\end{equation}
where $K_1 = k_{1f}/k_{1b} = c_2 c_3/c_1$ and $K_2 = k_{2f}/k_{2b} = c_3 c_4$ are the equilibrium constants for the primary hydrolysis and water dissociation reactions \cite{Baes1976}. Here, $k_{1b}$ and $k_{2b}$ serve as a penalty for departing from equilibrium; they must be large enough to enforce the equilibrium within the electrolyte but not to obstruct the simulation. $R_1$, $R_2$, $R_3$, and $R_4$ stand for the volumetric reaction rates associated with M$^{z_{1}+}$, M(OH)$^{(z_1 -1)+}$, H$^+$, and OH$^{-}$ ions. $R_5 = R_6 = 0$ as Na$^+$ and Cl$^-$ ions do not take part in the chemical reactions in Eqs. (\ref{eqn1}), (\ref{eqn2}) and (\ref{eqn3}).

The linear momentum balance equation for quasi-static loading with standard boundary conditions describes the deformation of the metal phase
\begin{equation} \label{eqn26}
\begin{aligned}
& \nabla \cdot \bm{\sigma} = \bm{0}  \quad\mathrm{} \text{in}\quad\Omega \\
\mathbf{t} = \mathbf{n}\cdot \bm{\sigma} = \mathbf{t}^0  \quad\mathrm{} &\text{on}\quad\partial\Omega_{\mathbf{t}} \quad\mathrm{}\quad \text{and} \quad\mathrm{}\quad \mathbf{u} = \mathbf{u}^0 \quad\mathrm{} \text{on} \quad\partial\Omega_{\mathbf{u}}
\end{aligned}
\end{equation}
where $\mathbf{t}^0$ and $\mathbf{u}^0$ are the prescribed traction and displacement vectors.

\begin{figure}[h!]
    \centering
    \includegraphics[width = 12cm]{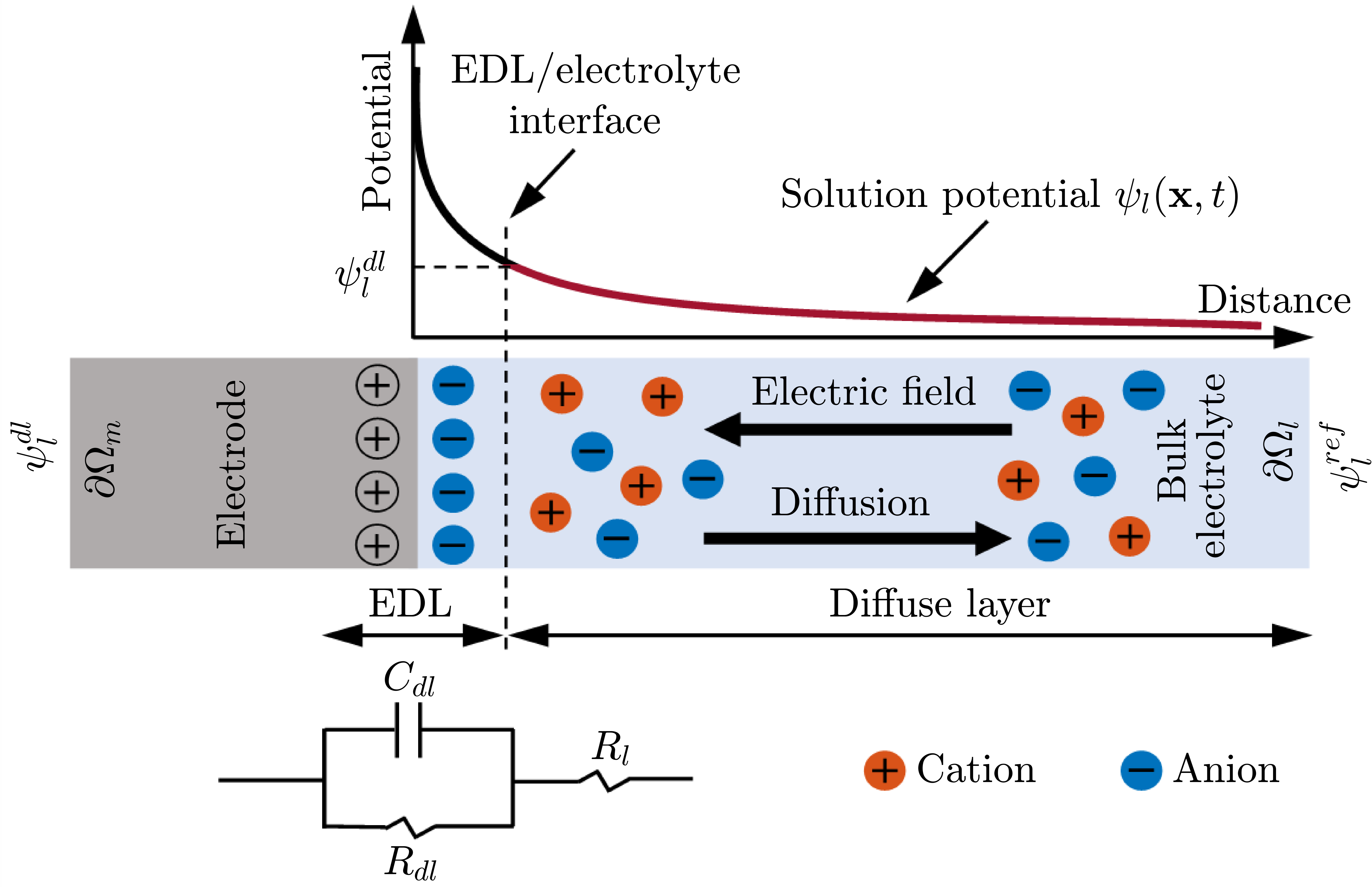}
    \captionsetup{labelfont = bf,justification = raggedright}
    \caption{Schematics of an electric double layer (EDL) and its electrode-electrolyte interface model with a resistor-capacitor equivalent circuit diagram.}
    \label{Fig2}
\end{figure}

The resulting set of governing equations is completed with the governing equation for the solution potential. The presence of a potential difference leads to the formation of an electric double layer (EDL) at the interface between the electrode and aqueous electrolyte, as shown in Fig. \ref{Fig2}. Ions in the electrolyte with the opposite charge than the electrode migrate and adsorb to the electrode surface due to the electrostatic forces \cite{Schmickler2020}. There is no motion of ions within the EDL. Beyond the EDL is the diffuse layer where ions are subjected to the combined effects of ion diffusion and electromigration \cite{Pilon2015}. The governing equation for the solution potential outside the EDL follows Ohm's law \cite{Jefimenko1989} along with the accompanying boundary conditions
\begin{equation}\label{eqn27}
\begin{aligned}
    & \nabla \cdot \lambda \nabla \psi_l = 0 \quad\mathrm{} \text{in}\quad\Omega \\
    \quad\mathrm{} \lambda \mathbf{n}\cdot \nabla \psi_l = 0 \quad\text{on}\quad\partial\Omega \quad &\psi_l = \psi_l^{dl} \quad\text{on}\quad\partial\Omega_m \quad\text{and} \quad \psi_l = \psi_l^{ref} \quad\text{on}\quad\partial\Omega_l
\end{aligned}
\end{equation}
where $\psi_l^{ref}$ is the solution potential on the reference electrode boundary $\partial\Omega_l$ and $\psi_l^{dl}$ is the potential at the interface between the EDL and the diffuse layer, Fig. \ref{Fig2}. The solution potential variable $\psi_l(\mathbf{x},t)$ is solved in the entire domain but only has physical meaning outside the EDL in the electrolyte domain. In Eq. (\ref{eqn27}), $\lambda = \lambda^s h(\phi) + (1-h(\phi)) \lambda^l$ is the effective conductivity interpolated between the conductivities in the liquid $\lambda^l$ and solid phases $\lambda^s$ ($\lambda^s \gg \lambda^l$). The electric conductivity in the electrolyte is expressed as a function of species concentrations: $\lambda_{l} = \sum_{i} D_{i}F^{2}z_{i}^{2}c_{i}/(RTV_{m})$ \cite{Neueder2014}. $\psi_l^{dl}$ is prescribed on the metal boundary $\partial\Omega_m$, Fig. \ref{Fig2}. High electrical conductivity in the metal phase $\lambda^s$ ensures a constant potential from the metal surface to the metal-electrolyte interface. A first-order resistor-capacitor equivalent circuit model is employed in this investigation to assess the evolution of the potential $\psi_l^{dl}$. The equivalent circuit model considered is schematically shown in Fig. \ref{Fig2}. Using this model, the evolution of the potential at the interface between the EDL and the electrolyte can be expressed as \cite{Sundararajan2020}    
\begin{equation}\label{eqn28}
\psi_l^{dl} = \psi^0 \Big( \frac{R_l}{R_l + R_{dl}} + \frac{R_{dl}}{R_l + R_{dl}} \text{exp}\Big( -\frac{t}{\tau_{dl}}\Big) \Big)
\end{equation}
where $\psi^0$ accounts for the initial surface polarization at time zero ($t=0$), $R_l$ the solution resistance, $R_{dl}$ the resistance of the EDL, $\tau_{dl} = R_{dl} C_{dl}$ the time constant that determines the rate at which the EDL is established, and $C_{dl}$ the capacitance of the EDL. For simplicity, it is assumed in the present work that the resistance of the EDL is proportional to the solution resistance: $R_{dl} = \chi R_l$, where $\chi \gg 1$ is the proportionality constant. It is further assumed that the capacitance of the EDL $C_{dl}$ is formed at half-time of capacitor charging $t_{c}$ and can be calculated as: $C_{dl} = \xi t_{c}/ (R_{dl} \text{ln}2)$ \cite{Sundararajan2020}, where $\xi$ is the geometric factor introduced to account for the change in the electrode-electrolyte interfacial area during the dissolution process, taken to be $\xi = 1$ for planar electrode-electrolyte surfaces \cite{Pilon2015}. The phase-field variable $\phi$, evaluated at $\phi = 1/2$, is used to track the change in the interfacial area. Upon these assumptions, the evolution of the potential at the interface between the EDL and the diffuse layer can be written as
\begin{equation}\label{eqn29}
\psi_l^{dl} = \psi^0 \Big(\frac{1}{1 + \chi} + \frac{\chi}{1 + \chi} \text{exp} \Big( -\frac{t}{\xi t_c} \big) \Big)
\end{equation}
The equivalent circuit parameters $\chi$ and $t_c$ are calibrated against experimental measurements (Section \ref{sec3}). It is emphasized that the previous expression accounts for the EDL without explicitly simulating it in the computational domain. To the best of our knowledge, it is the first time the EDL charging effect is considered in a corrosion damage-prediction model.

Alternative approaches to solve for the electric potential in the electrolyte in the context of phase-field corrosion models are typically based on an interface propagation rate source term that supplements Eq. ({\ref{eqn27}}) \cite{Chadwick2018, Ansari2018, LIN2019, LIN2021, Chen2022, Tsuyuki2018, CUI2023, Brewick2022}. In the case of sharp interface models \cite{SUN2014, Duddu2014, Brewick2017, Brewick2019}, the electric potential distribution is determined by solving Eq. ({\ref{eqn27}}) and prescribing current density at the metal-environment interface obtained from the Tafel or Butler-Volmer equations \cite{{Jones1996}}. However, these equations fail to capture the transient in surface polarisation due to EDL charging since they describe steady-state dissolution. This effect will cause discrepancies between the current density measurement and modeling output when the measurement times are similar to the EDL charging time. By utilizing the equivalent circuit approximation (Fig. {\ref{Fig2}}) to describe the EDL charging behavior, the present model is the first to consider this effect in a transient context, suitable for both damage prediction during accelerated corrosion tests and for stress corrosion cracking predictions over arbitrary time scales. The possibilities given by the present approach are tested thoroughly in Section {\ref{sec3}} and Section {\ref{sec4}}.

\subsection{Mechano-electro-chemical coupling} \label{sec24}

\begin{figure}[h!]
    \centering
    \includegraphics[width = 16 cm]{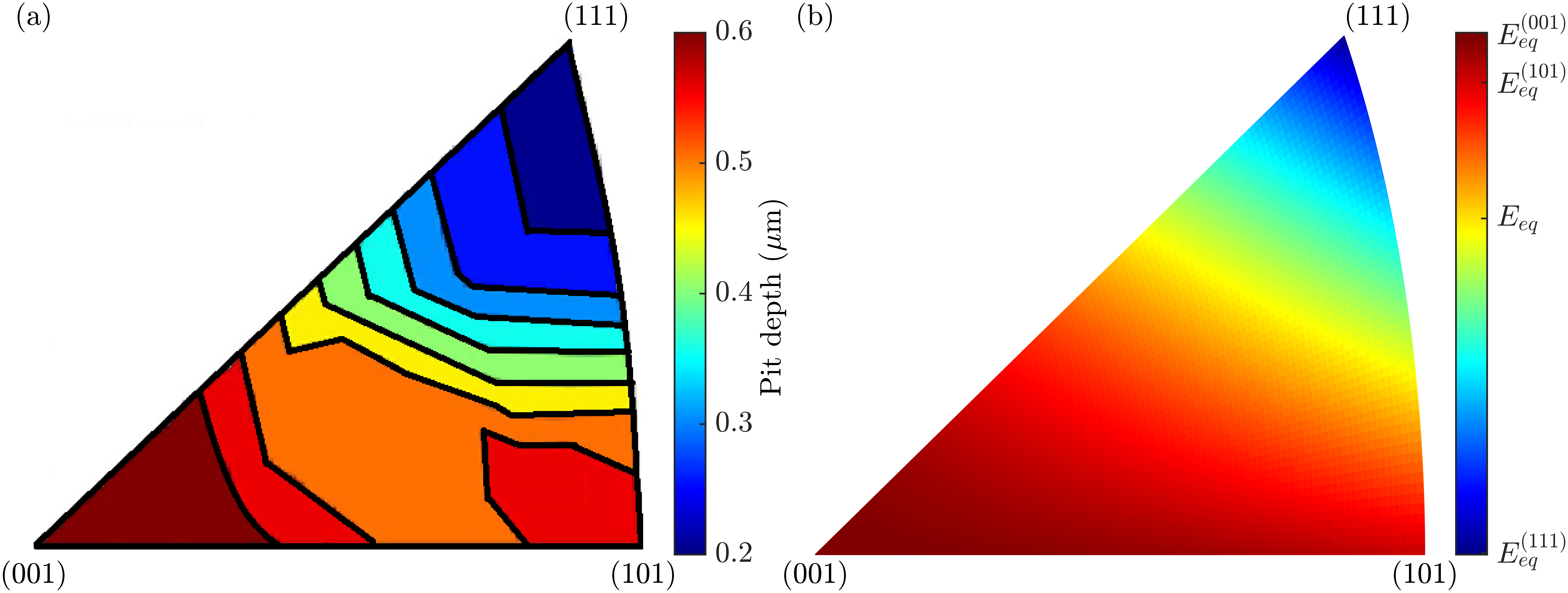}
    \captionsetup{labelfont = bf,justification = raggedright}
    \caption{Dependence of corrosion potential on crystallographic orientation. (a) Experimental measurements of pit depth in stainless steel as a function of grain orientation \cite{Lindell2015}. (b) Directional variation in equilibrium corrosion potential determined based on the experimental measurements in (a). $E_{eq}$ stands for the macroscopic equilibrium corrosion potential. The polarisation variation map in (b) is derived by solving three Tafel equations, one for each corner, subjected to a known applied potential and a zero net polarisation variation upon the summation. (the polycrystalline value).}
    \label{Fig3}
\end{figure}

The rate of the electrode reaction (flux) is proportional to current density. Following the usual notation in the literature, this can be written as $v = i_a / (zF)$, where $v$ is the rate of reaction, $i_a$ is the current density, and $z$ is the number of electrons in the electrochemical reaction. Following the commonly used phenomenological Butler-Volmer model, the current density is expressed as \cite{Jones1996,cui2022generalised}
\begin{equation} \label{eqn30}
\begin{aligned}
i_a^\theta = i_0 \Big[\text{exp}\Big (\frac{\alpha z_1 F \eta^{\theta}}{RT}\Big) - \text{exp}\Big(-\frac{(1 - \alpha) z_1 F \eta^{\theta}}{RT}\Big)\Big]
\end{aligned}
\end{equation}
where $i_0$ is the exchange current density, $\alpha$ is the anodic charge transfer coefficient ($\alpha = 0.26$ in this work \cite{CUI2023}), and $\eta^{\theta}$ is the crystallographic-orientation dependent overpotential defined as
\begin{equation}\label{eqn31}
    \eta^{\theta} = E_{app} - E_{eq}^{\theta}-\psi_{l} \quad\mathrm{}\quad E_{eq}^{\theta} = E_{eq} + \Delta E^{\theta}
\end{equation}
where $E_{app}$ is the applied electric potential and $E^{\theta}_{eq}$ is the equilibrium corrosion potential that depends on grain orientation. It is assumed that $E^{\theta}_{eq}$ deviates from the \textit{macroscopic} equilibrium corrosion potential $E_{eq}$ for a certain value that depends on crystallographic orientation $\Delta E^{\theta}$. The equilibrium corrosion potential is determined on the microstructural level within a single grain, while macroscopic equilibrium corrosion potential is obtained on the macroscale level considering all exposed grains. Thus, the macroscopic corrosion potential is approximately an average of the crystallographic-dependent corrosion potentials exhibited by single grains. The variation in equilibrium corrosion potential as a function of grain orientation is defined based on experimental measurements for pitting corrosion in stainless steel \cite{Lindell2015}. As shown in Fig. \ref{Fig3}(a), pitting is less pronounced in the (111) plane compared to the (001) and (101) planes \cite{Lindell2015}. Based upon these measurements for pit depth and simplifying the corrosion kinetics with the Tafel equation \cite{Uhlig2008}, the dependence of corrosion potential on crystallographic orientation is obtained and depicted in Fig. \ref{Fig3}(b). The generated equilibrium corrosion potential map is asymmetric with respect to the macroscopic equilibrium corrosion potential $E_{eq}$ as follows from the experimental data \cite{Lindell2015}. The maximum deviation from the macroscopic equilibrium corrosion potential $\Delta E^{\theta}_{max}$ is represented as the difference between the equilibrium corrosion potentials of the (111) and (001) planes: $\Delta E^{\theta}_{max} = E^{(111)}_{eq} - E^{(001)}_{eq}$. The average value of the corrosion potentials of the (111), (001), and (101) planes returns the macroscopic equilibrium corrosion potential $E_{eq}$. The effect of $\Delta E^{\theta}_{max}$ on the corrosion behavior in polycrystalline materials is examined in Section \ref{sec4}.

The role of mechanical fields in accelerating corrosion kinetics is incorporated by following the work by Cui et al. \cite{CUI2023}, which is based on Gutman's mechanochemical theory \cite{Gutman1988}. Accordingly, the electrode kinetics is represented by an extended Butler-Volmer model
\begin{equation} \label{eqn32}
                   i = i_a^\theta \Big(\frac{\varepsilon^{p}}{\varepsilon_{0}}+1 \Big)\exp\Big(\frac{\sigma_h V_m}{RT}\Big)
\end{equation}
where $i$ is the local current density of the mechanically deformed metal, $i_a^\theta$ is the current density of the non-deformed electrode given in Eq. (\ref{eqn30}) and $\sigma_{h}$ is the hydrostatic stress. The effect of crystallographic orientation on the mechanical properties is expressed by transforming the mechanical properties from the principal material directions into the system coordinates, Eq. (\ref{eqn18}). 

The electrode kinetics are governed by the local current density $i$ (Eq. (\ref{eqn32})) while the phase-field mobility parameter $L$ (Eq. (\ref{eqn21})) controls the motion of the solid-liquid interface. Using Eq. (\ref{eqn32}) and considering the linear relation between the phase-field mobility $L$ and the current density \cite{MAI2016, CUI2021,cui2022generalised}, an analogous expression for the mobility parameter can be obtained as
\begin{equation}\label{eqn33}
        L = L_0 \Big[\text{exp}\Big (\frac{\alpha z_1 F \eta^{\theta}}{RT}\Big) - \text{exp}\Big(-\frac{(1 - \alpha) z_1 F \eta^{\theta}}{RT}\Big)\Big] \Big(\frac{\varepsilon^{p}}{\varepsilon_{y}}+1 \Big)\exp\Big(\frac{\sigma_h V_m}{RT}\Big)
\end{equation}
where $L_0$ is the mobility parameter corresponding to the exchange current density $i_0$ in the absence of mechanical fields. The kinematic parameter $L_0$ is calibrated against experimental measurements of pit depth in the following section.

\subsection{Numerical implementation} \label{sec25}

The governing equations are implemented in the commercial finite element software COMSOL Multiphysics\footnote{The code developed will be available at \url{www.imperial.ac.uk/mechanics-materials/codes} after article acceptance.} \cite{Comsol}. The computational domain of the metal and the corrosive environment is discretized using triangular finite elements with second-order Lagrangian interpolation functions. The characteristic element size is selected to be at least four to five times smaller than the interface thickness $\ell$ to capture a smooth transition between the metal and the environment. This proved to be sufficient based on previous works \cite{CUI2021,cui2022generalised, CUI2023, KOVACEVIC2023}. The mesh is only refined in the vicinity of the metal-liquid interface and the expected area of interface propagation. A relatively coarse mesh is kept far away from the interface. The size of the refined mesh area is selected to be large enough to ensure the interface stays within it at all times. An implicit time-stepping method is used for temporal discretization. A monolithic (fully coupled) solution algorithm is adopted for solving the governing equations. The solver accuracy in each time step is controlled by a relative tolerance of 10$^{-4}$. The maximum time increment is constrained to $\Delta t$ = 0.2 s in the simulation for model calibration (Section \ref{sec3}). Further decreases in the time step and mesh size in a convergence analysis (Section A.1.) show no impact on the results, Fig. A1. The same ratio between the time step and the final computational time is kept in the remaining simulations in Section \ref{sec4}. 

\section{Model calibration and validation} \label{sec3}

\begin{figure}[h!]
    \centering
    \includegraphics[height = 7 cm]{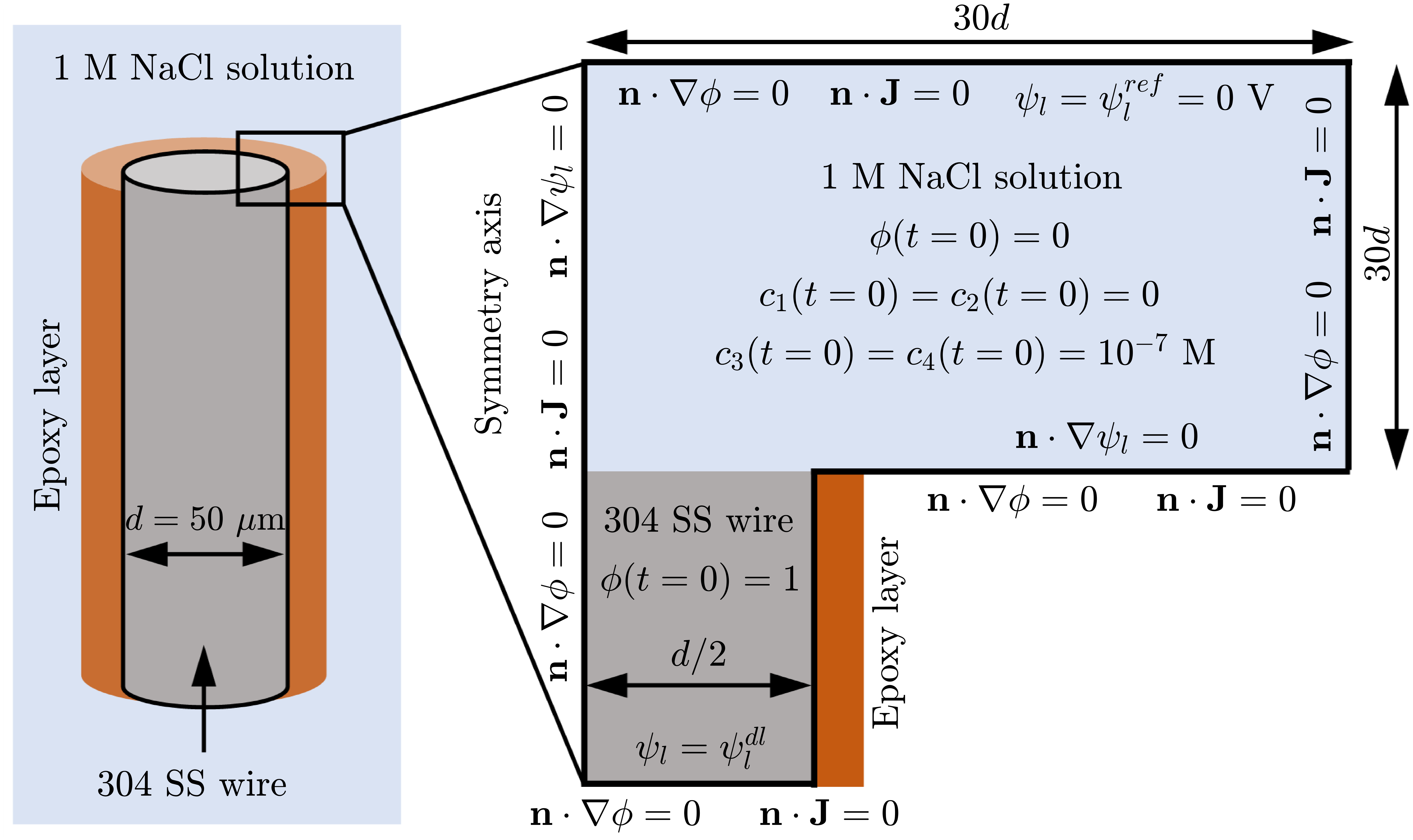}
    \captionsetup{labelfont = bf,justification = raggedright}
    \caption{Schematic disposition of the experimental setup used in Ref. \cite{ERNST2002a} (left) and the corresponding computational domain (right) for the 304 stainless steel (SS) wire immersed in 1 M NaCl solution.}
    \label{Fig4}
\end{figure}

The experimental setup given in Ref. \cite{ERNST2002a} and schematically illustrated in Fig. \ref{Fig4} is used to calibrate and validate the present model. A 304 stainless steel (SS) wire with a 50 $\mu$m diameter was immersed in 1 M NaCl solution. The wire was circumferentially coated in epoxy resin such that only the cross section of the wire was exposed to the corrosive environment. The experiment was conducted in potentiostatic loading. The current was supplied to the system until the potential difference between the wire and a reference electrode was 600 mV, measured against a saturated calomel electrode (vs. SCE). During the experiment, the current was controlled to maintain this potential difference. Experimental pit depth measurements and current density as a function of immersion time are used to calibrate model parameters. The model parameters are the kinematic mobility coefficient $L_0$ (Eq. ({\ref{eqn33}})) that controls the motion of the interface and the equivalent circuit parameters $\chi$ and $t_c$ (Eq. ({\ref{eqn29}})) that describe the ratio between the EDL resistance and bulk resistance of the electrolyte and the half-time of capacitor charging.

The numerical simulations are performed by considering an axisymmetric domain as depicted in Fig. \ref{Fig4}. The electrolyte size is significantly larger than the metal to replicate the experimental setup in Ref. \cite{ERNST2002a} and ensure that the solution does not saturate with metal ions, as this would stop the process. The governing equations (\ref{eqn21}), (\ref{eqn22}), and (\ref{eqn27}) are solved with accompanying initial and boundary conditions. No flux boundary condition for diffusion and phase-field is imposed at the outer boundaries of the electrolyte to simulate an unbounded computational domain. The same condition is prescribed along the exterior metal surface to mimic the epoxy layer. These boundary conditions imply that no diffusion occurs across the domain boundary. Zero reference solution potential is enforced on the far top boundary in the electrolyte. The potential at the interface between the EDL and the electrolyte $\psi_l^{dl}$, obtained using Eq. (\ref{eqn29}), is imposed on the far bottom metal surface. High metal electrical conductivity ensures a constant potential from the bottom metal surface to the metal-electrolyte interface. As there is no change in the electrode-electrolyte interfacial area, the geometric factor in Eq. (\ref{eqn29}) is set to $\xi = 1$. The remaining boundaries are treated as perfectly isolated with no flux boundary condition for the solution potential. A pH of 7 is used to calculate the initial concentrations of H$^+$ and OH$^-$ ions in the electrolyte domain. The initial concentration of Na$^+$ and Cl$^-$ ions in solution follows those used in the experiment. The initial concentration of metal ions $\text{M}^{z_1+}$ in the solid phase is equated to the site density of the material. No $\text{M}^{z_1+}$ and $\text{M}(\text{OH})^{(z_1-1)+}$ ions are initially present in the liquid phase. In this simulation, the mechanical effect and the variation in corrosion potential are not considered ($E_{eq}^{\theta} = E_{eq}$).

The material properties of the alloy, the diffusivity of ions in the electrolyte, and the equilibrium reaction constants $K_{1}$ and $K_{2}$ used in the simulation are obtained from the literature and listed in Table \ref{table1}. The backward reaction constants $k_{1b} = $ 200 m$^3$/(mol$\cdot$s) and $k_{2b} = $ 1000 m$^3$/(mol$\cdot$s) are chosen to be sufficiently large to enforce the equilibrium conditions (Eq. (\ref{eqn25})) but not so large to induce convergence issues. The equilibrium concentration in the liquid phase $c_1^{l,eq}$ is defined based on the solubility of salts formed on the exposed metal surface \cite{Isaacs1995} as the formation of corrosion products and the protective layer are neglected in the present model, Eqs. (\ref{eqn1}), (\ref{eqn2}), (\ref{eqn3}). The \textit{macroscopic} equilibrium corrosion potential $E_{eq}$ and the equilibrium concentration in the solid phase $c_1^{s,eq}$ are determined using the molar fraction of alloying elements \cite{BS2005}. The charge number $z_1$ of metal ions is obtained in the same fashion. Charge numbers of the other ions have common values, as indicated in Eqs. ({\ref{eqn2}}) and ({\ref{eqn3})}. The phase-field parameters $\kappa$ and $\omega$ are linked to the interfacial energy $\Gamma$ and the interface thickness $\ell$, Eq. (\ref{eqn12}). The interface thickness $\ell$ is set to be significantly smaller than the diameter of the stainless steel wire ($\ell$ = 5 $\mu$m). The average value of the interfacial energy reported for stainless steel in Ref. \cite{Kanhaiya2014} is used in this investigation. The chemical free energy density curvature parameter $A$ in Eq. (\ref{eqn8}) is assumed to have a similar value as in Refs. \cite{MAI2016, CUI2021, KOVACEVIC2023}.

\begin{table} [h!]
\centering
\captionsetup{labelfont = bf,justification = raggedright}
\caption{Parameters common to all phase-field simulations.}
\begin{tabular}{l l l} 
\hline
Quantity & Value & Unit\\
\hline
Equilibrium concentration in the solid phase $c_1^{s,eq}=1/V_m$ & 144.3 & mol/L \cite{BS2005}\\   
Equilibrium concentration in the liquid phase $c_1^{l,eq}$ & 5.1 & mol/L \cite{Isaacs1995}\\
Diffusion coefficient of M$^{z_1+}$ in the liquid phase $D^l_{1}$ & $0.719$ $\times$ $10^{-9}$ & m/s$^2$ \cite{Haynes2016}\\
Diffusion coefficient of M(OH)$^{(z_1-1)+}$ in the liquid phase $D^l_{2}$ & $0.719$ $\times$ $10^{-9}$ & m/s$^2$ \cite{Haynes2016}\\
Diffusion coefficient of H$^{+}$ in the liquid phase $D^l_{3}$ & $9.311$ $\times$ $10^{-9}$ & m/s$^2$ \cite{Haynes2016}\\
Diffusion coefficient of OH$^{-}$ in the liquid phase $D^l_{4}$ & $5.273$ $\times$ $10^{-9}$ & m/s$^2$ \cite{Haynes2016}\\
Diffusion coefficient of Na$^{+}$ in the liquid phase $D^l_{5}$ & $1.334$ $\times$ $10^{-9}$ & m/s$^2$ \cite{Haynes2016}\\
Diffusion coefficient of Cl$^{-}$ in the liquid phase $D^l_{6}$ & $2.032$ $\times$ $10^{-9}$ & m/s$^2$ \cite{Haynes2016}\\
Interfacial energy $\Gamma$ & 2.10 & J/m$^2$ \cite{Kanhaiya2014}\\
Chemical free energy density curvature parameter $A$ & $1.02$ $\times$ $10^8$ & J/m$^3$ \cite{CUI2021, KOVACEVIC2023}\\
Primary hydrolysis equilibrium constant $K_{1}$ & $3.1622$ $\times$ $10^{-7}$ & mol/m$^3$ \cite{Baes1976}\\
Water dissociation equilibrium constant $K_{2}$ & $10^{-8}$ & mol$^2$/m$^6$ \cite{Baes1976}\\
Macroscopic equilibrium corrosion potential $E_{eq}$ & -0.729 & V (vs. SCE) \cite{BS2005}\\
Solid phase conductivity $\lambda_{s}$ & 10$^6$ & S/m \cite{BS2005}\\
\hline
\end{tabular}
\label{table1}
\end{table}

\begin{figure}[h!]
    \centering
    \includegraphics[width = 16cm]{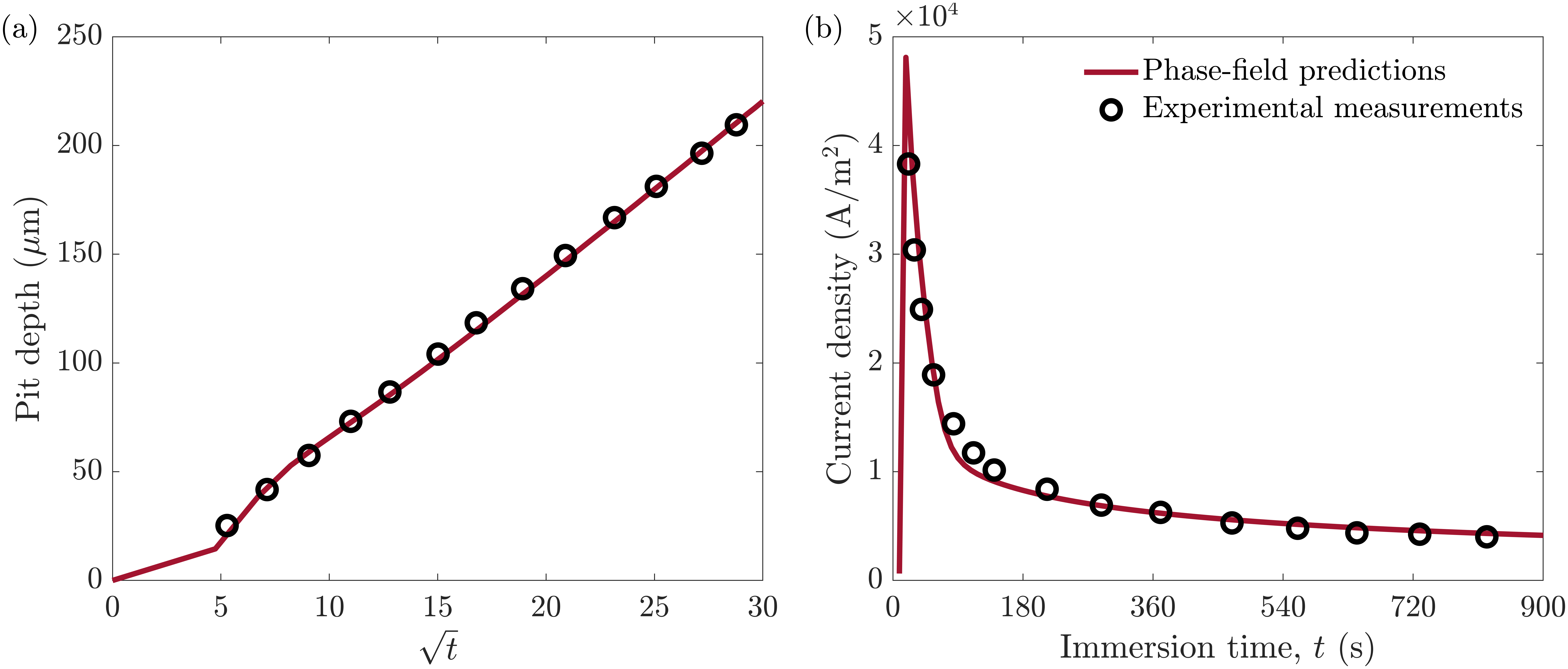}
    \captionsetup{labelfont = bf,justification = raggedright}
    \caption{Comparison between experimental measurements \cite{ERNST2002a} and phase-field predictions of (a) the evolution of the pit depth and (b) current density as a function of immersion time in 1 M NaCl solution.}
    \label{Fig5}
\end{figure}

The predicted pit depth and current density are plotted as a function of immersion time in Fig. \ref{Fig5}, together with experimental measurements from Ref. \cite{ERNST2002a}. The phase-field results and the experimental data are in good agreement. Before the EDL charges up, the initial pit kinetics is in an activation-controlled regime as shown in Fig. \ref{Fig5}(a) at early times. The interface reactions govern the material transport. The metal is not yet affected by the electrostatic field of adsorbed and migrated ions. The solution potential almost entirely cancels the electrode polarization, significantly reducing the overpotential in Eq. (\ref{eqn31}). Afterward, there is a short transition from the activation to the diffusion-controlled stage when the electric potential drop at the EDL builds up enough for the metal ions to release. As the process proceeds, the pit growth kinetics becomes diffusion-controlled. Long-range interactions and the motion of ions far away from the interface dictate the corrosion rate at that stage. Diffusion and electromigration govern the transport of ions in the electrolyte. Before the system reached the diffusion-controlled process the pit depth was not measured in the experiment \cite{ERNST2002a}. However, extrapolating the experimental measurements in Fig. \ref{Fig5}(a) indicates that the pit kinetics was not in a diffusion-controlled regime at early times, which agrees with the phase-field predictions.

The results obtained for the current density are presented in Fig. \ref{Fig5}(b). The model predictions agree well with the experimental data. During the charging of the EDL, the current density increases rapidly until the EDL is sufficiently charged. At this point, the current starts decreasing and stabilizes at longer times. The width and height of the peak in the current density response at early times are determined by the capacitor charging half-time $t_c$ and the resistance of the EDL (i.e., the constant of proportionality $\chi$ in the present model), Eq. (\ref{eqn29}). At longer times $t \gg t_{c}$, the current density response and the total migration of metal ions are controlled by the resistance of the EDL. In this case, where dissolution times are significantly larger than $t_c$, the potential at the interface between the EDL and the electrolyte is governed by the first term in Eq. (\ref{eqn29}). This implies that while both constants are necessary to capture the experimental response in accelerated corrosion tests and short dissolution times, the resistance of the EDL is only needed to capture natural dissolution (activation-controlled corrosion). The slight discrepancy between the experimental data and the phase-field predictions at early times in Fig. \ref{Fig5}(b) may result from oversimplifying the equivalent circuit model by assuming one capacitive process. A more complex equivalent circuit model would add more capacitors, and thus, more constants to fit the width of the current density decay peak.

The good agreement between the experiment and the simulation indicates that the present model is accurately calibrated in terms of the interfacial mobility $L_0$ (Eq. (\ref{eqn33})) and the equivalent circuit model parameters $\chi$ and $t_c$ in Eq. (\ref{eqn29}). The constructed model qualitatively reproduces the experimental data using an interfacial mobility parameter $L_0 = 1.2$ $\times$ $10^{-15}$ m$^3$/(J$\cdot$s), a half-time of capacitor charging $t_c = 10$ s, and a resistance proportionality constant $\chi = 120$. The values for $t_c$ and $\chi$ agree well with experimental works that have used the same equivalent circuit model (Fig. \ref{Fig2}) for the same type of stainless steel and a similar aqueous environment \cite{Dong2013, Vogiatzis2016}. It should be noted that models in the literature \mbox{\cite{Chadwick2018, Ansari2018, Tantratian2022}} are validated using only measurements of pit depth \mbox{\cite{ERNST2002a}}. However, the proposed framework can capture measurements of pit depth and current density.

\begin{figure}[h!]
    \centering
    \includegraphics[height = 6.5cm]{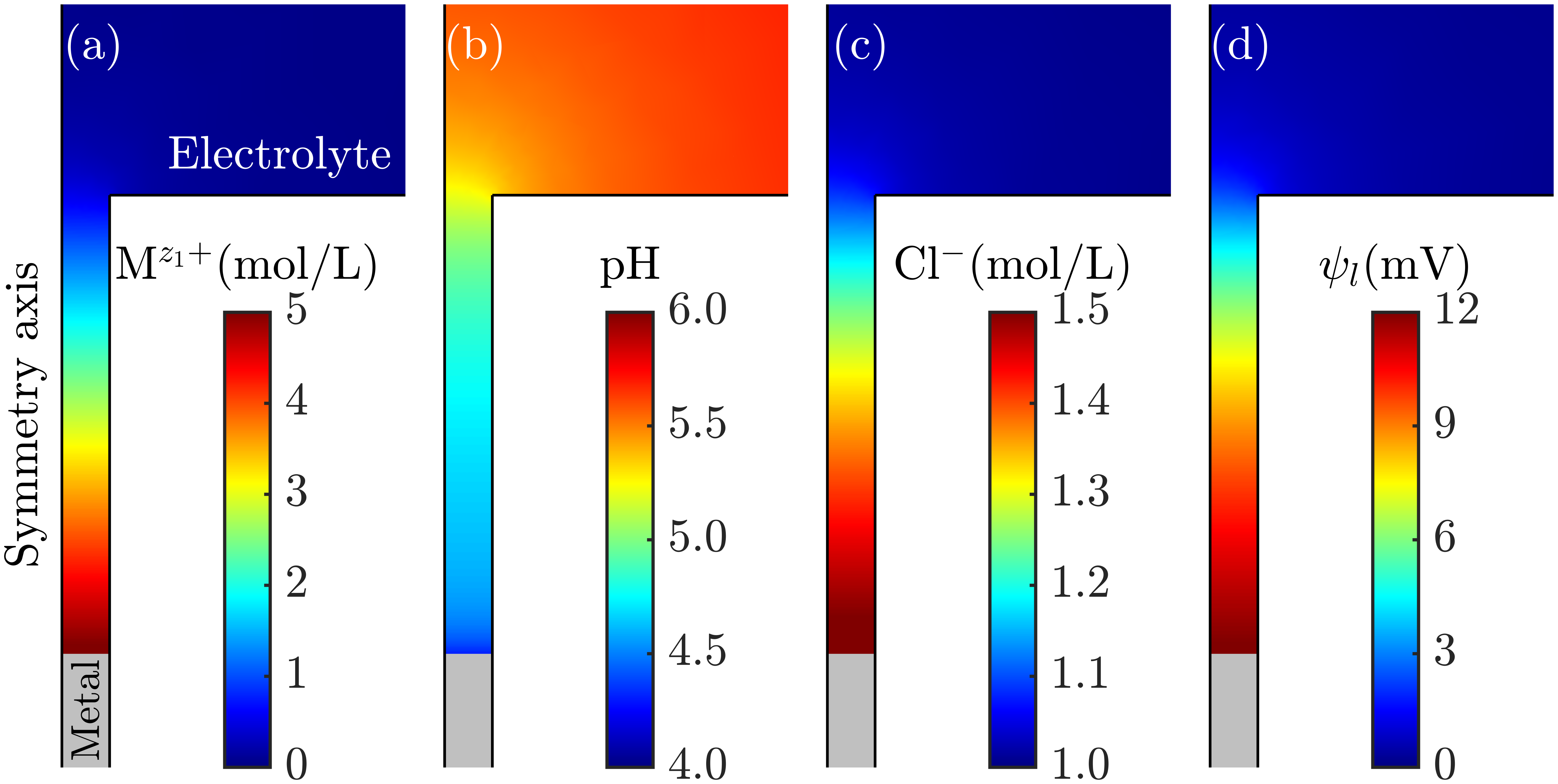}
     \captionsetup{labelfont = bf,justification = raggedright}
    \caption{Distribution of (a) metal ions $\text{M}^{z_1+}$, (b) pH, (c) chloride ions Cl$^-$, and (d) solution potential at the final computational time. The legend bars apply to the whole computational domain.}
    \label{Fig6}
\end{figure}

\begin{figure}[h!]
    \centering
    \includegraphics[width = 16cm]{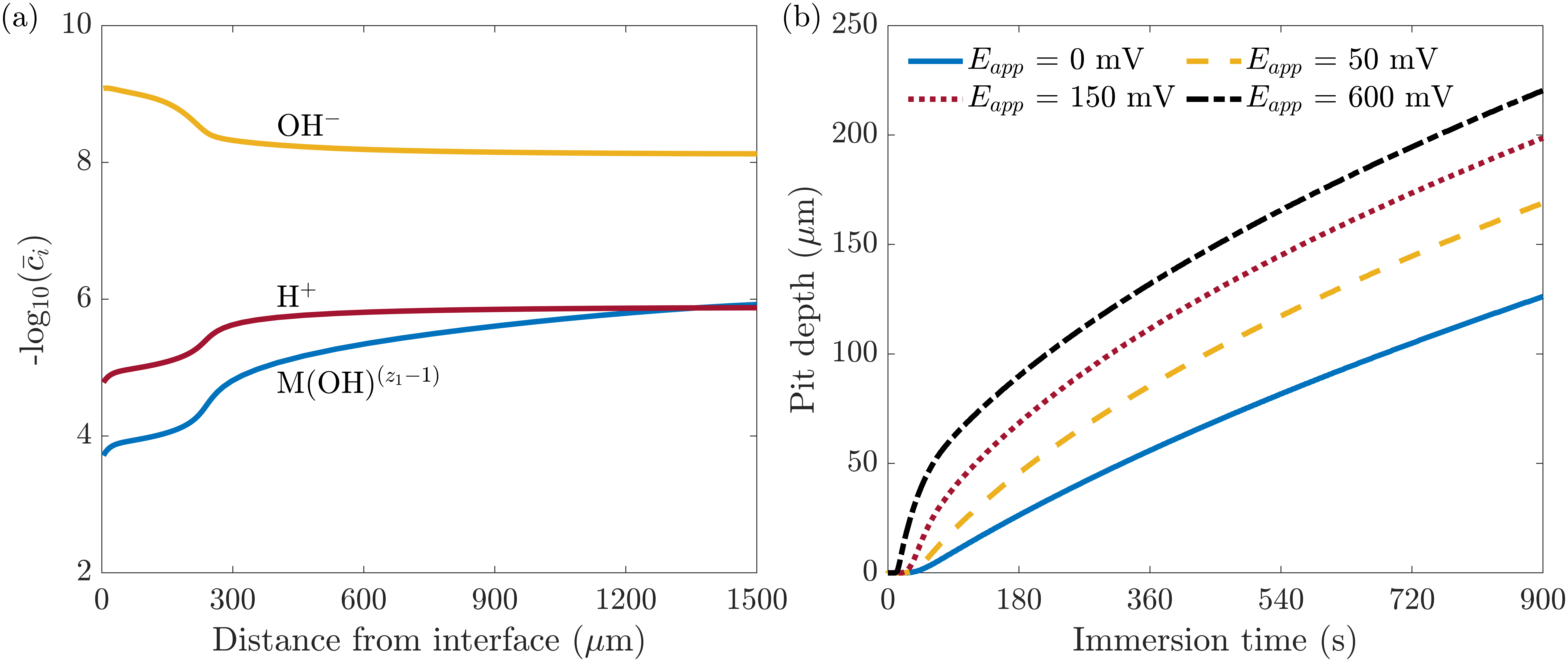}
    \captionsetup{labelfont = bf,justification = raggedright}
    \caption{Electrochemical predictions. (a) Concentration profile of M(OH)$^{(z_1-1)+}$, H$^{+}$, and OH$^{-}$ as a function of distance from the interface at the final computational time for 600 mV (vs. SCE). Concentration values $\bar{c}_i$ are normalized with 1 mol/L. (b) Pit depth as a function of immersion time for different applied potentials. The potentials are expressed vs saturated calomel electrode.}
    \label{Fig7}
\end{figure}

The distribution of concentration of representative ions and solution potential at the final computational time is shown in Fig. \ref{Fig6}. As depicted in Fig. \ref{Fig6}(a), the metal ion concentration gradually reaches the equilibrium concentration in the liquid phase ($c_1^{l,eq}$) at the interface, indicating that the process is in a diffusion-controlled stage. This agrees with the pit kinetics in Fig. \ref{Fig5}(a). The production of H$^+$ ions in the primary hydrolysis reaction (Eq. ({\ref{eqn2}))} reduces the pH of the solution near the interface, as indicated in Fig. \ref{Fig6}(b). The overall acidity does not exceed a pH of 6, which is consistent with the literature \cite{Uhlig2008, Miyamoto2018, Miyamoto2022}. Consequently, the concentration of OH$^-$ drops to satisfy the water dissociation equilibrium, which in turn raises the concentration of M(OH)$^{(z_1-1)+}$ as this is the second product of the hydrolysis reaction. The concentration profile of H$^+$, OH$^-$, and M(OH)$^{(z_1-1)+}$ ions along the electrolyte at the final computational time is given in Fig. \ref{Fig7}(a). As can be observed in Fig. \ref{Fig7}(a), the sum of pH and pOH is 14 throughout the entire domain. This confirms that the backward reaction constants $k_{1b}$ and $k_{2b}$ in Eq. (\ref{eqn23}) are adequately chosen to satisfy the equilibrium conditions. The Cl$^-$ ion concentration distribution is given in Fig. \ref{Fig6}(c). Although the simulation starts with a uniform concentration of Cl$^-$ ions and with no contribution to the chemical reactions (Eqs. (\ref{eqn1}), (\ref{eqn2}), and (\ref{eqn3})), Cl$^-$ ions move to the metal-liquid interface due to electromigration as being negatively charged. An opposite behavior is noticed for Na$^+$ ions (plot not shown here) as they are positively charged. As indicated in Fig. \ref{Fig6}(d), the solution potential gradient dominates near the interface and slowly diminishes in the bulk electrolyte. Far from the interface where the solution potential is not strong enough, Cl$^-$ and Na$^+$ ions are uniformly distributed. 

The sensitivity of the model to externally applied potentials is illustrated in Fig. \ref{Fig7}(b) for various applied potentials: 600 mV, 150 mV, 50 mV, and 0 mV (vs. SCE). The latter corresponds to an activation-controlled regime. The material properties and the model parameters used in the sensitivity analysis are the same as in the previous simulation. As depicted in Fig. \ref{Fig7}(b), the model naturally captures the transition from activation- to diffusion-controlled regimes by applying different potentials to the system. At low applied potentials, the pit kinetics remain in an activation-controlled regime. As the applied potential increases, a transition from activation- to diffusion-controlled regimes occurs. A further increase in applied potential leads to diffusion-controlled corrosion and faster pit kinetics. The results obtained in Fig. \ref{Fig7}(b) are in agreement with the literature \cite{TIAN2015}. The effects of varying the applied potential from activation- to diffusion-controlled processes on the corrosion behavior in polycrystalline materials are discussed in Section \ref{sec4}.

The agreement between the experimental measurements and phase-field predictions of pit depth and current density indicates that the present model can accurately simulate uniform corrosion. The model satisfactorily predicts pit kinetics and electrochemical measurements and exhibits sensitivity to externally applied potentials. In the following section, the model is exploited to ascertain the role of microstructure on pitting and stress corrosion cracking in polycrystalline materials.

\section{Microstructure-sensitive phase-field
simulations and discussion} \label{sec4}

Two case studies are considered to assess the evolution of corrosion in polycrystalline materials. The first one examines pitting corrosion associated with the local rupture of a protective layer in a polycrystalline material with various grain morphology distributions. The second one involves a polycrystalline material loaded in tension with an initial defect that acts as a stress concentrator, leading to the formation of pit-to-crack transition and crack propagation.

\subsection{Pitting corrosion} \label{sec41}

\subsubsection{Pitting corrosion method} \label{sec411}
Synthetically generated microstructures with an average grain size of 20 $\mu$m, 40 $\mu$m, and 60 $\mu$m are considered in this work. Ten different microstructures are randomly created for each grain size to obtain a statistically significant sample. In total, thirty different grain morphology distributions are generated. These are given in the Appendix in Figs. A2, A3, and A4. Two values of $\Delta E^{\theta}_{max}$ are considered to investigate the influence of the degree of variation in corrosion potential. For each set of ten microstructures with specified grain size, the following values of $\Delta E^{\theta}_{max}$ are considered: $\Delta E^{\theta}_{max} = 10\%E_{eq} = 73$ mV (vs. SCE) and $\Delta E^{\theta}_{max} = 5\% E_{eq} = 36.50$ mV (vs. SCE), returning a total of sixty simulations. Each simulation is subjected to applied potentials of 600 mV and $-479$ mV (vs. SCE) to capture the diffusion and activation-controlled processes. The former corresponds to experiments used in Section \ref{sec3} for accelerated corrosion. The latter ensures the natural dissolution of metals with an overpotential of 250 mV \cite{Uhlig2008}. Hence, one hundred twenty simulations are carried out to analyze pitting corrosion in a polycrystalline material. A special case with no variation in corrosion potential $\Delta E^{\theta}_{max} = 0$ mV (vs. SCE) is also simulated. This case corresponds to the homogeneous material without the effect of microstructure and is a reference study for comparison. In this simulation, the equilibrium corrosion potential is set to $E_{eq}^{\theta} = E_{eq}$. The mechanical contribution is not considered in this set of simulations for pitting corrosion.

\begin{figure}[h!]
    \centering
    \includegraphics[height = 7.0 cm]{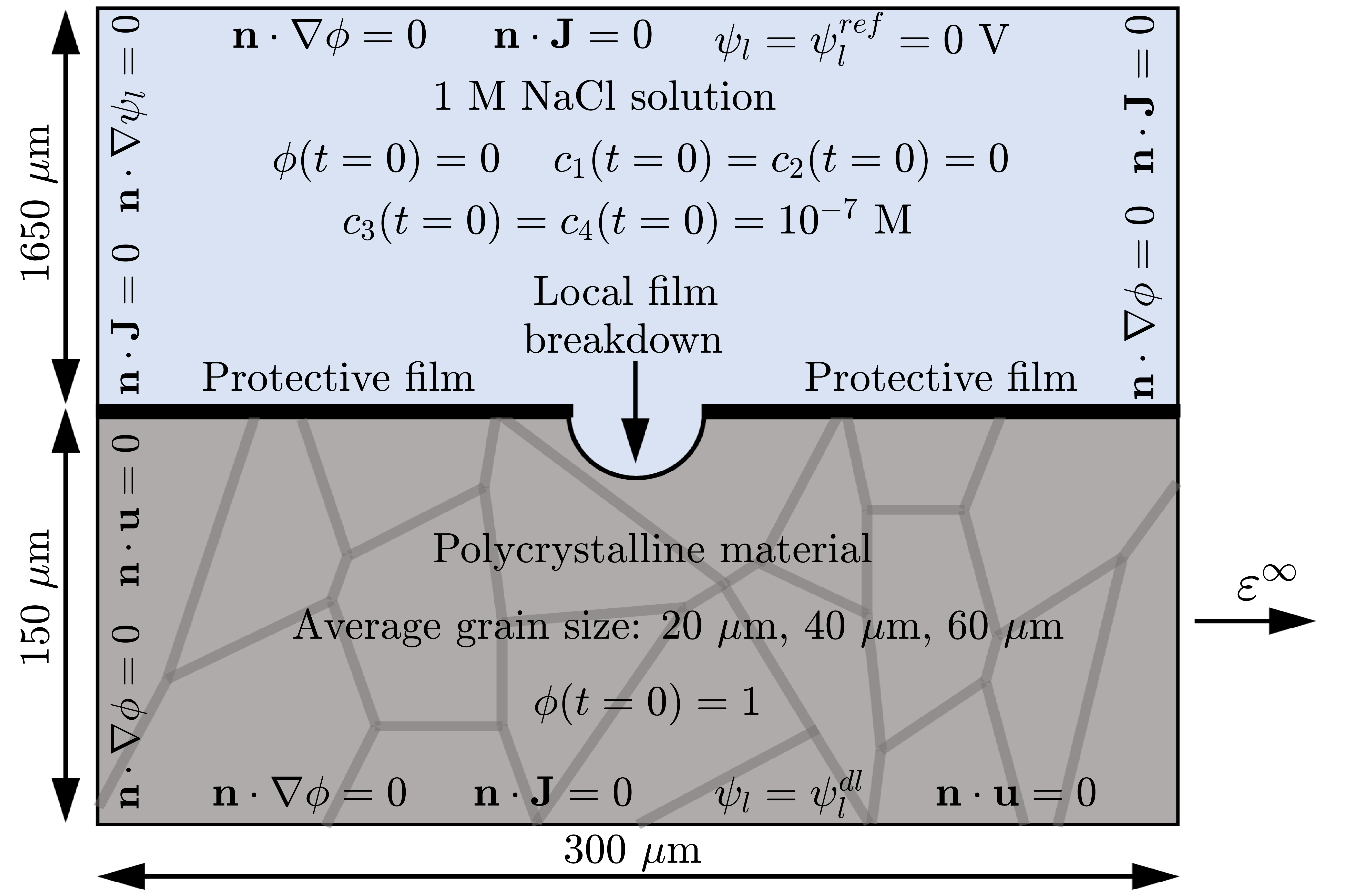}
    \captionsetup{labelfont = bf,justification = raggedright}
    \caption{Computational domain consisting of the polycrystalline material and the surrounding corrosive environment. The semi-circular initial defect created by the rupture of the protective layer has a radius of 6 $\mu$m.}
    \label{Fig8}
\end{figure}

The computational domain includes the polycrystalline material and the surrounding electrolyte, as shown in Fig. \ref{Fig8}. The material is assumed to be a 304 stainless steel alloy and the corrosive environment is a 1 M NaCl solution. The size of the electrolyte is significantly larger than the material to avoid saturation effects. A thin protective layer is assumed on the metal surface. The layer is locally damaged in a small area, which acts as a nucleation site for pitting corrosion. The initial defect is semi-circular with a radius of 6 $\mu$m. The initial concentration of the ionic species follows those in the previous example (Section \ref{sec3}). No flux boundary conditions are enforced at all the outer boundaries of the computational domain for diffusion and phase-field. The protective layer on the metal surface is modeled as an impermeable barrier with a thickness of 1 $\mu$m, along with the corresponding no flux boundary condition for all variables. The far top boundary in the electrolyte represents the reference electrode with a prescribed zero potential. The potential at the interface between the EDL and the electrolyte $\psi_l^{dl}$ (Eq. (\ref{eqn29})) is applied to the bottom boundary in the metal. The phase-field variable at $\phi = 1/2$ is used to track the evolution of the electrode-electrolyte interfacial area and determine the geometric factor in Eq. (\ref{eqn29}). The remaining boundaries hold no flux condition for the solution potential.

The Euler angles $\varphi_1$, $\varphi_2$, and $\varphi_3$ are randomly assigned to each grain. The equilibrium corrosion potential of each grain is determined by using Eq. (\ref{eqn31}). $\Delta E^{\theta}$ in Eq. (\ref{eqn31}) is obtained using the Euler angles of each grain. The other material properties of the metal and the electrolyte follow those in the previous example in Section \ref{sec3}, Table 1. The same interface thickness of $\ell$ = 5 $\mu$m and the phase-field parameters used in Section \ref{sec3} are employed in these simulations. The kinetic parameter $L_0$ and the equivalent circuit model parameters $\chi$ and $t_c$ are chosen on previously determined comparison with experimental data.

\subsubsection{Pitting corrosion results} \label{sec412}
\begin{figure}[h!]
    \centering
    \includegraphics[width=15 cm]{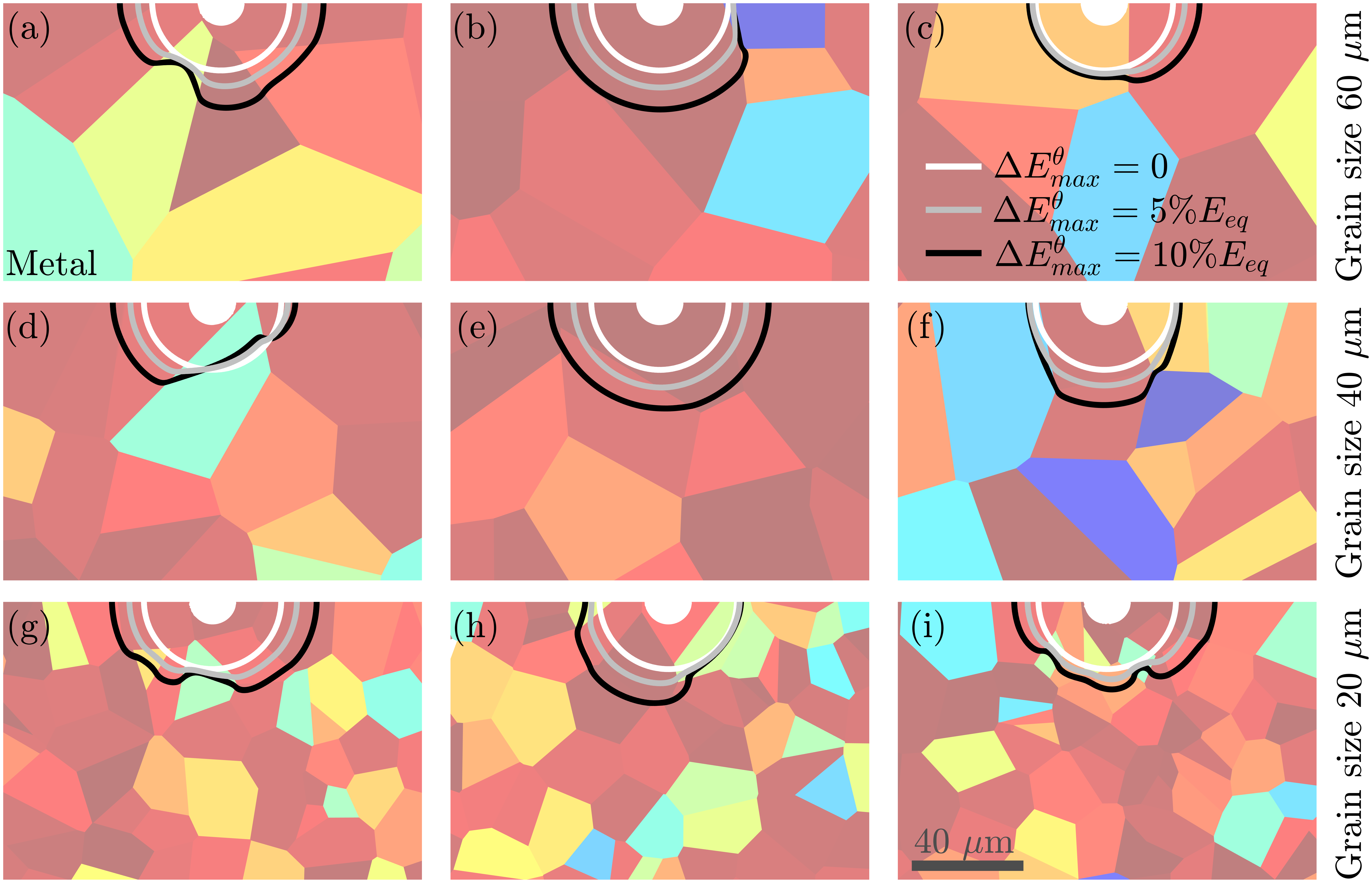}
    \captionsetup{labelfont = bf,justification = raggedright}
    \caption{Evolved pit shapes at the final computational time for activation-controlled corrosion for three representative microstructures with an average grain size of (a-c) 60 $\mu$m, (d-f) 40 $\mu$m, and (g-i) 20 $\mu$m. The black and gray lines represent pit shapes for nonhomogenous equilibrium corrosion potential with $\Delta E^{\theta}_{max} = 10\% E_{eq}$ and $\Delta E^{\theta}_{max} = 5\% E_{eq}$ variations. The white line stands for pitting corrosion with a uniform corrosion potential. The surrounding environment is not shown in the plots. The scale bar applies to each plot. The color scale for variations in equilibrium corrosion potential is given in Fig. \ref{Fig3}.}
    \label{Fig9}
\end{figure}

The results of the phase-field simulations for activation-controlled corrosion with nonhomogenous and uniform equilibrium corrosion potentials are given in Fig. \ref{Fig9}. Three representative microstructures are shown for each grain size considered. The results show that the variation in corrosion potential promotes nonuniform pit growth and the formation of facets, resulting in sharp changes in pit shape. This finding is consistent with the computational \cite{Sahu2022, SAHU2023} and experimental literature \cite{Koroleva2007, Yasuda1990}. The higher the variation in corrosion potential, the larger the pit area. Moreover, the larger grain size yields more severe damage compared to the fine microstructure with a grain size of 20 $\mu$m. Especially if the pitting front meets unfavourable crystallographic orientations for anodic dissolution, like close-packed planes. This highlights that the variations in equilibrium corrosion potential and grain size strongly affect the pit shape and dissolved metal area. The role of nonuniform pit shapes in the presence of mechanical loading in stress and strain distributions is discussed in the following section.

On the other hand, a constant corrosion potential retains uniform pit growth, maintaining the initial semicircular pit shapes at all times. The pit area is smaller than in the previous cases with nonhomogenous equilibrium corrosion potentials, which is in agreement with previous studies \cite{Brewick2017, Brewick2019}. The results in Fig. \ref{Fig9} show that microstructures with unfavorable crystallographic orientations may lead to severe damage compared to the homogeneous case. Employing homogeneous materials without considering the effect of microstructure may lead to an underestimation of localized damage. A more quantitative comparison between these cases with nonhomogenous and uniform equilibrium corrosion potential is addressed in Section \ref{sec43}.

The obtained results for diffusion-controlled corrosion, not given here, indicate that the pitting kinetics and pit shape remain unaffected by variations in equilibrium corrosion potential, and thus, grain size. The pit shape is the same under nonhomogenous and uniform equilibrium corrosion potential, which is in agreement with expectations as the process is controlled by diffusion far away from the metal-electrolyte interface. Regardless of the corrosion potential variation or grain size, the pit growth remains uniform, keeping the initial semicircular pit shapes at all times. Therefore, considering homogeneous and nonhomogeneous materials in accelerated corrosion (diffusion-controlled) returns the same estimation of damage and dissolved metal area. The fact that diffusion-controlled dissolution is invariant of crystallographic anisotropy has been observed experimentally \cite{ERNST2002a, TIAN2015} and it is the basis for metal surface electropolishing \cite{COELHO2020}.

\subsection{Stress-assisted corrosion} \label{sec42}

\subsubsection{Stress-assisted corrosion method} \label{421}
In these simulations, the material is simultaneously immersed in a corrosive environment and subjected to tensile strain. The computational domain corresponds to Fig. \ref{Fig8}. Thirty microstructures generated in the previous section with an average grain size of 20 $\mu$m, 40 $\mu$m, and 60 $\mu$m (ten microstructures for each grain size) are considered in this case study. Three different variations in equilibrium corrosion potential are applied to each microstructure: $\Delta E^{\theta}_{max} = 10\% E_{eq} = 73$ mV (vs. SCE), $\Delta E^{\theta}_{max} = 5\% E_{eq} = 36.50$ mV (vs. SCE), and $\Delta E^{\theta}_{max} = 0$ mV (vs. SCE). The latter represents a scenario where the material has anisotropic mechanical properties but uniform corrosion potential. We also consider a special case that stands for the homogeneous material without the effect of microstructure. This special case serves as a reference study for comparison. It should be noted that the present study considers stress-assisted corrosion, as described in Eqs. (32) and (33). Following previous works \cite{CUI2021,cui2022generalised, Brewick2022}, the present model can be extended to incorporate the effect of hydrogen embrittlement, the film rupture-dissolution-repassivation mechanism, and mechanical rupture.

The simulations are conducted under an applied potential of $-479$ mV (vs. SCE) to avoid diffusion-controlled effects. The initial and boundary conditions for diffusion, phase-field, and solution potential are the same as those used in Fig. \ref{Fig8}. Additional boundary conditions are prescribed for the mechanical equilibrium equation. The normal component of the displacement vector along the left and bottom boundary is constrained ($\mathbf{n} \cdot \mathbf{u} = 0$). A non-zero remote uniaxial tensile strain $\varepsilon^{\infty}$ is enforced on the right edge, as shown in Fig. \ref{Fig8}. The magnitude of the remote strain is varied to examine the role of mechanical loading in enhancing corrosion kinetics.

\begin{figure}[h!]
    \centering
    \includegraphics[width = 16cm]{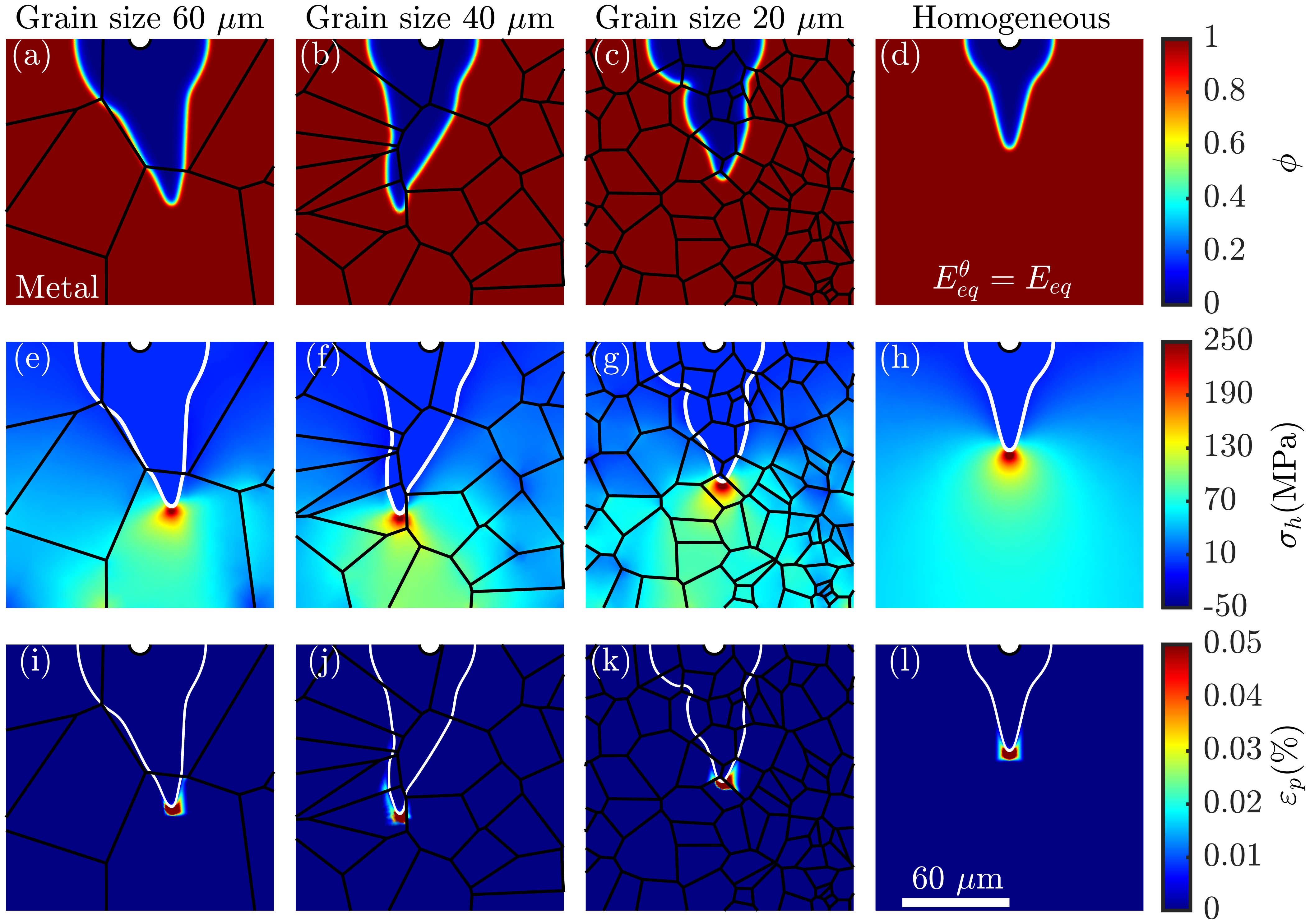}
    \captionsetup{labelfont = bf,justification = raggedright}
    \caption{Contour plots of (a-d) the phase-field variable, (e-h) hydrostatic stress $\sigma_h$, and (i-l) the effective plastic strain $\varepsilon^p$ for microstructures with an average grain size of 20 $\mu$m, 40 $\mu$m, 60 $\mu$m, and homogeneous material at the final computational time. $\Delta E^{\theta}_{max} = 10\% E_{eq}$ variation in equilibrium corrosion potentials and remote uniaxial tensile strain $\varepsilon^{\infty} = 0.06\%$ are used in the plots. The scale bar and color scale apply to each plot.}
    \label{Fig10}
\end{figure}

The same material properties for the metal, electrolyte, and phase-field parameters are used in this case study as in the previous example on pitting corrosion. The equivalent circuit parameters $\chi$ and $t_c$ also follow those in the previous study. The influence of mechanical fields on the interface kinetics is incorporated via Eq. (\ref{eqn33}). The kinetic parameter $L_0$ is obtained previously through comparison with experiments (Section \ref{sec3}). The mechanical properties of the stainless steel from the literature are employed in the simulation. The three monocrystal elastic constants are $C_{11}$ = 209.0 GPa, $C_{12}$ = 133.0 GPa, and $C_{44}$ = 121.0 GPa \cite{Teklu2004}. The mechanical properties of the alloy are transformed into the global coordinate system using the transformation matrix $\mathbf{R}$ defined based on the Euler angles of each grain, Eq. (\ref{eqn18}). For the homogeneous material without the effect of microstructure, the material behavior is characterized by Young's modulus of 199 GPa and Poisson's ratio of 0.29 \cite{Teklu2004}. To describe plastic deformation, the von Mises yield criterion is used with power-law isotropic hardening (Eq. (\ref{eqn20})). The yield stress of 205 MPa, the yield strain $\varepsilon_0 = 1$ $\times$ $10^{-3}$, and the strain hardening exponent $N = 0.2$ are used in the simulation \cite{deOliveira2019}.

\subsubsection{Stress-assisted corrosion results} \label{422}
The obtained results in terms of phase-field contours and mechanical fields are given in Fig. \ref{Fig10}. The representative microstructure is shown for each grain size considered, along with the homogeneous material without the effect of microstructure. The results consider a variation in equilibrium corrosion potential of 73 mV (vs. SCE) and remote uniaxial elastic tensile strain $\varepsilon^{\infty} = 0.06\%$. Fig. \ref{Fig10} indicates that the underlying microstructure with dependencies of corrosion potential and mechanical properties on crystallographic orientation plays a significant role in determining the growth of defects, defect (crack) morphology, and the distribution of hydrostatic stress and equivalent plastic strain. The microstructural features lead to more extensive defects, irregular crack shapes, and highly nonuniform hydrostatic stress and plastic strain distributions. In contrast, the homogeneous material without the effect of microstructure yields smaller defects and exhibits uniform hydrostatic stress and plastic strain distributions, yielding symmetric crack morphology.

\begin{figure}[h!]
    \centering
    \includegraphics[width = 16cm]{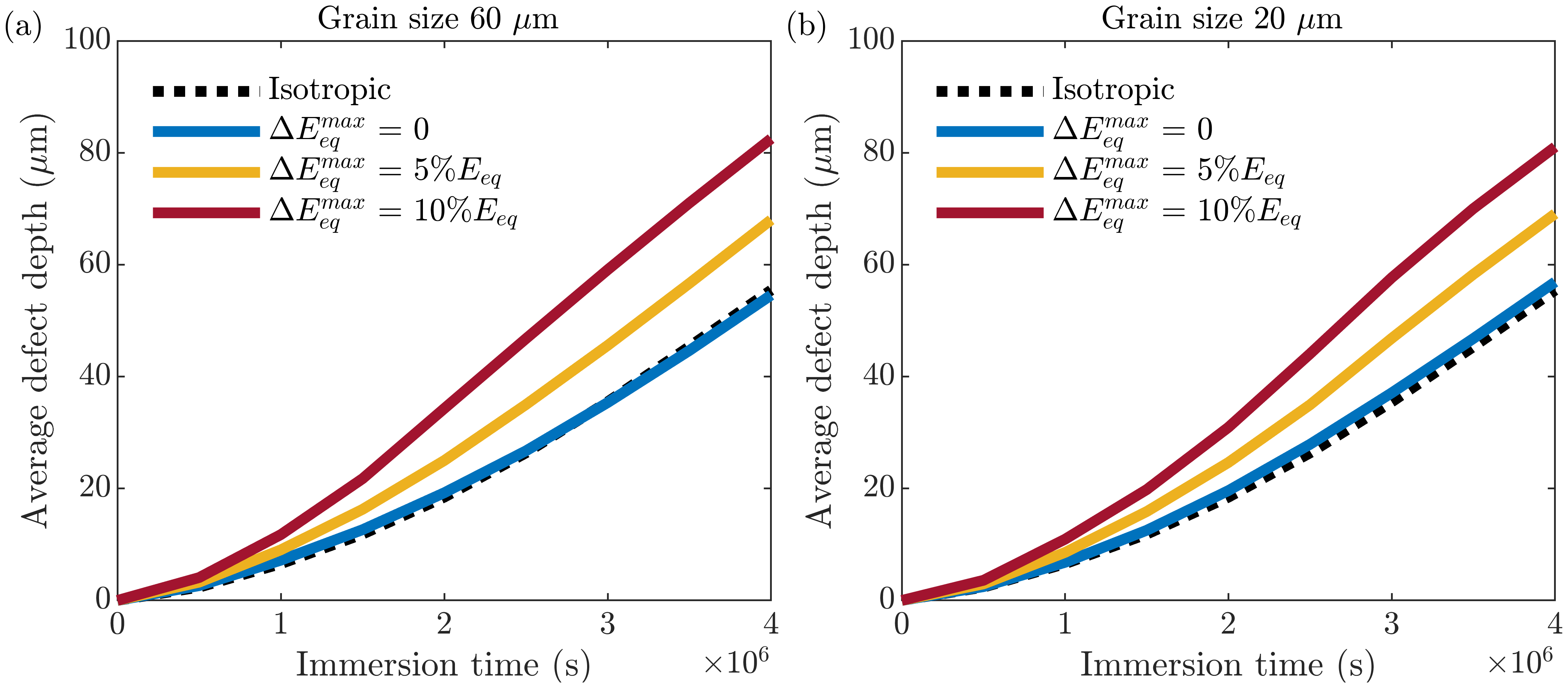}
    \captionsetup{labelfont = bf,justification = raggedright}
    \caption{Average defect growth as a function of immersion time for different variations in equilibrium corrosion potential $\Delta E^{\theta}_{max}$ and microstructures with an average grain size of (a) 60 $\mu$m and (b) 20 $\mu$m under uniaxial tensile strain $\varepsilon^{\infty} = 0.06\%$.} 
    \label{Fig11}
\end{figure}

\begin{figure}[h!]
    \centering
    \includegraphics[width = 15 cm]{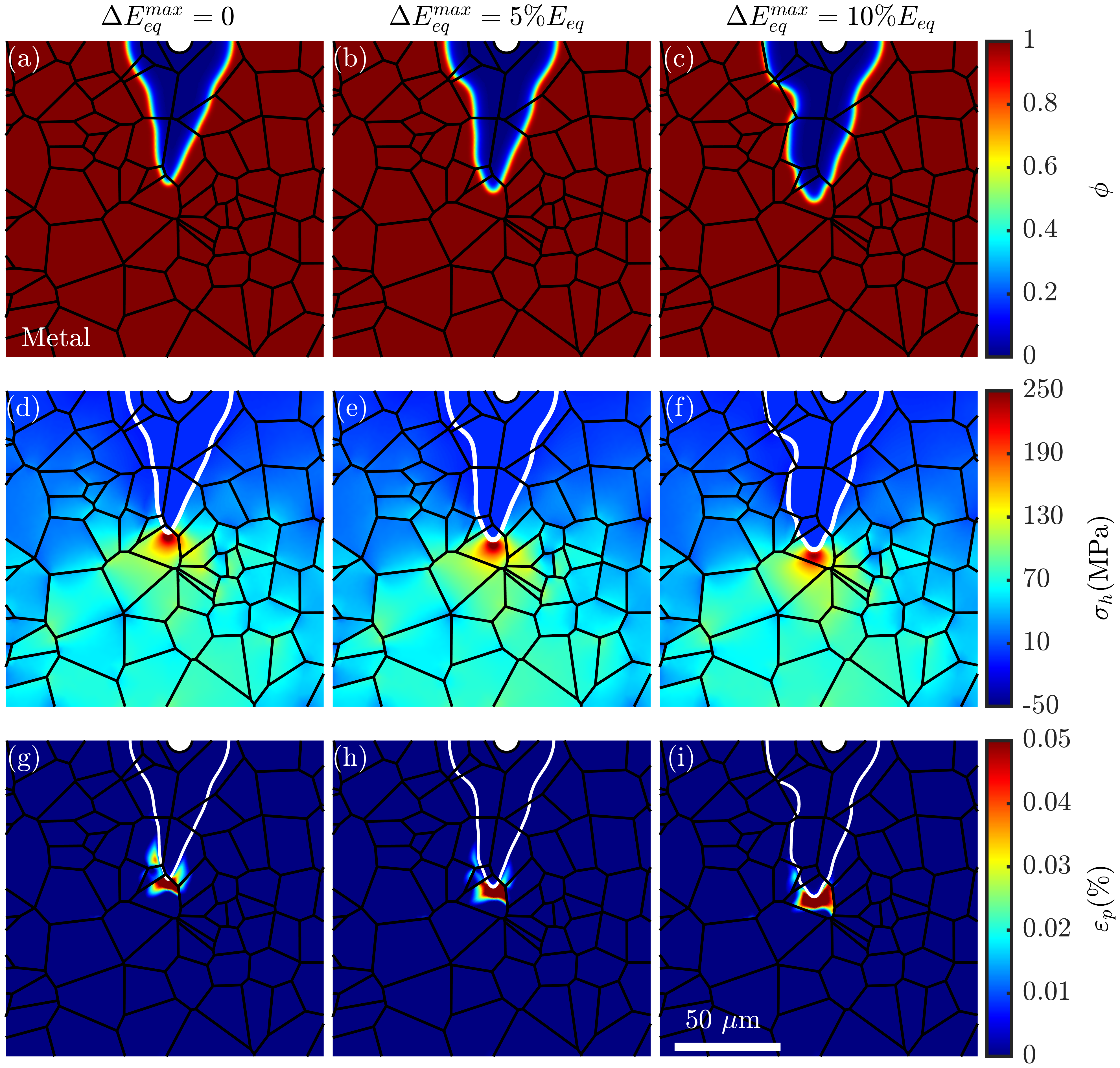}
    \captionsetup{labelfont = bf,justification = raggedright}
    \caption{Contour plots of (a-c) the phase-field variable, (d-f) hydrostatic stress $\sigma_h$, and (g-i) the effective plastic strain $\varepsilon^p$ for microstructures with an average grain size of 20 $\mu$m subjected to uniaxial tensile strain $\varepsilon^{\infty} = 0.06\%$ and different variations in equilibrium corrosion potential. The scale bar and color scale apply to each plot.} 
    \label{Fig12}
\end{figure}

The role of variation in equilibrium corrosion potential in the growth of defects is presented in Fig. \ref{Fig11}. The comparison is made between microstructures with an average grain size of 20 $\mu$m and 60 $\mu$m under remote uniaxial tensile strain $\varepsilon^{\infty} = 0.06\%$. The defect kinetics is represented as the average value obtained from ten simulations for each grain size. The results reveal that defect growth increases with an increase in the variation in equilibrium corrosion potential for both grain sizes. Similar kinetics are returned by comparing the microstructures with an average grain size of 20 $\mu$m and 60 $\mu$m. When compared to a scenario with uniform corrosion potential, the defect kinetics are significantly altered, resulting in a more rapid crack growth. The results in Fig. \ref{Fig11} also demonstrate that the anisotropic mechanical properties with uniform corrosion potential yield the same defect kinetics as that of a homogeneous material without the effect of microstructure.

In addition to altering the defect kinetics, the variation in equilibrium corrosion potential also influences the crack morphology, as shown in Fig. \ref{Fig12} at the final computational time. The representative microstructure with an average grain size of 20 $\mu$m subjected to uniaxial tensile strain $\varepsilon^{\infty} = 0.06\%$ is considered. Fig. \ref{Fig12} indicates that increasing the variation in equilibrium corrosion potential leads to more irregular crack shapes. The variation in corrosion potential and mismatch in elastic constants produce nonhomogenous hydrostatic stress and plastic strain distributions, which govern the shape of the evolving defect through mechanical coupling in Eq. (\ref{eqn33}). Although anisotropic mechanical properties with uniform corrosion potential do not affect the defect kinetics when compared to the homogeneous material (Fig. \ref{Fig11}), they cause nonuniform hydrostatic stress (Fig. \ref{Fig12}(d)) and plastic strain distributions (Fig. \ref{Fig12}(g)) with high local values in the vicinity of GBs. Consequently, higher maximum stresses are returned than in the case of a homogeneous material, Fig. \ref{Fig10}. This nonuniform distribution of mechanical fields changes the crack path and promotes asymmetrical crack morphologies, Fig. \ref{Fig12}(a). In the case of a homogeneous material without the effect of microstructure, uniform hydrostatic stress and plastic strain distributions with symmetric crack morphologies are obtained, Fig. \ref{Fig10}. The present results suggest that the underlying microstructure plays a role in determining the hydrostatic stresses and plastic strains that dictate the shape of the evolving defect.    

\begin{figure}[h!]
    \centering
    \includegraphics[width = 15 cm]{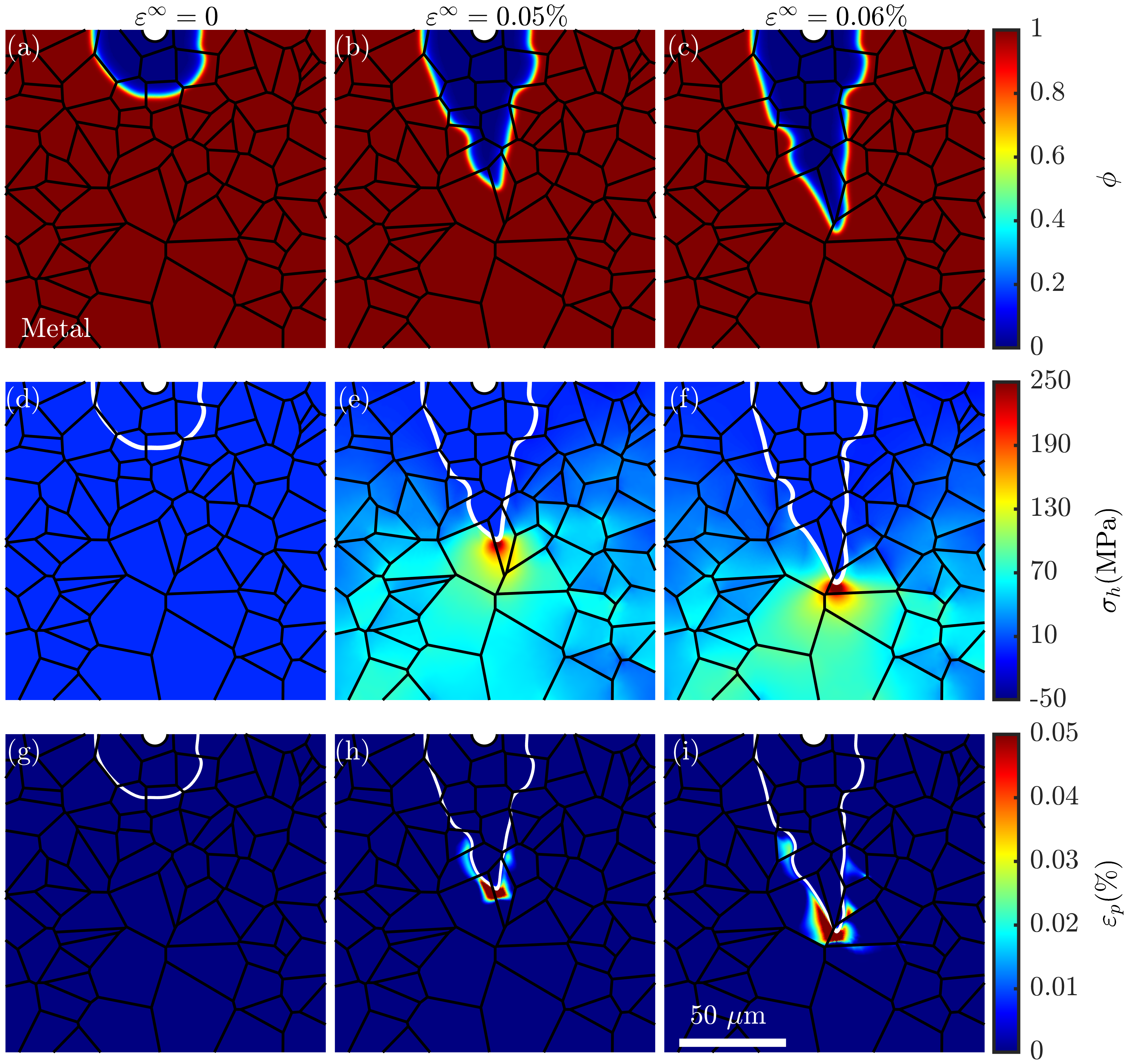}
    \captionsetup{labelfont = bf,justification = raggedright}
    \caption{Contour plots of (a-c) the phase-field variable, (d-f) hydrostatic stress $\sigma_h$, and (g-i) the effective plastic strain $\varepsilon^p$ for microstructures with an average grain size of 20 $\mu$m for different applied uniaxial tensile strains. $10\% E_{eq}$ variation in equilibrium corrosion potentials is used in the plots. The scale bar and color scale apply to each plot.} 
    \label{Fig13}
\end{figure}

The effect of mechanical loading in enhancing corrosion kinetics is demonstrated in Fig. \ref{Fig13}. The microstructure with an average grain size of 20 $\mu$m and $10\% E_{eq}$ variation in equilibrium corrosion potential are considered in Fig. \ref{Fig13}. In the absence of mechanical loading ($\varepsilon^{\infty} = 0$), irregular pit shapes occur due to anisotropic dissolution, as previously shown in Fig. \ref{Fig9}. The application of a uniaxial tensile strain ($\varepsilon^{\infty} = 0.05\%$) produces stresses and plastic strains in a localized area. As the pit evolves, the hydrostatic stress and plastic strain increase and contribute to the interface kinetics (Eq. (\ref{eqn33})). As the process proceeds, the mechanical fields change the pit morphology and initiate a pit-to-crack transition, which ultimately leads to crack propagation at longer times. In this case, hydrostatic stresses predominantly govern the growth of the defect, as the plastic strain is relatively low. A further increase in the applied uniaxial tensile strain ($\varepsilon^{\infty} = 0.06\%$) yields higher hydrostatic stresses engaged in a more extensive area and triggers noticeable plastic deformations, even at early times. Both hydrostatic stress and plastic strain contribute to the interface kinetics and govern the shape of the evolving defect. As a result, longer cracks are obtained compared to the previous case ($\varepsilon^{\infty} = 0.05\%$). In the presence of mechanical straining irregular crack morphologies occur due to the anisotropic mechanical properties and equilibrium corrosion potential.

\begin{figure}[h!]
\centering
\includegraphics[height = 7.4cm]{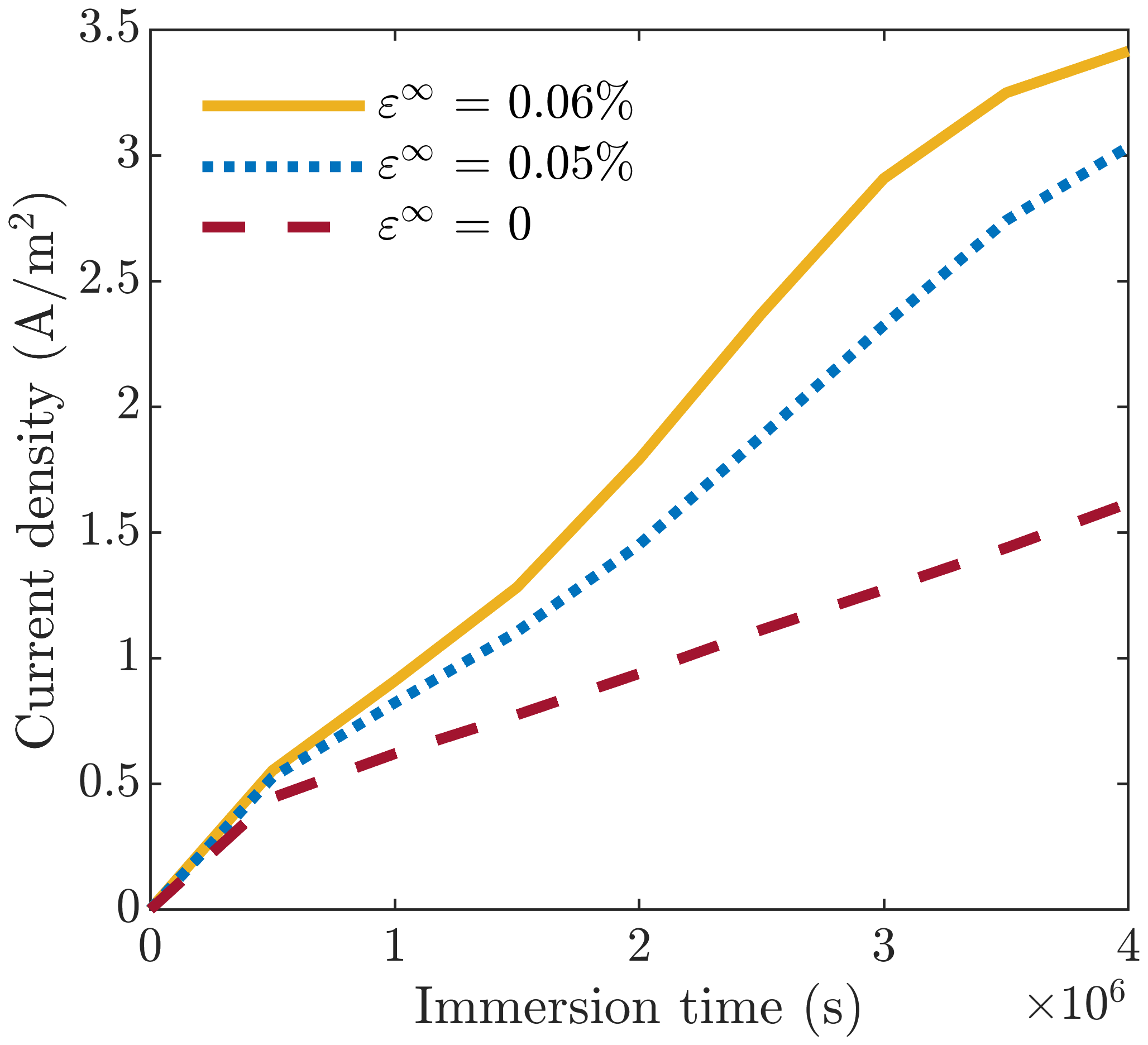}
 \captionsetup{labelfont = bf,justification = raggedright}
\caption{Current density as a function of immersion time for different applied uniaxial tensile strains.}
\label{Fig14}
\end{figure}

Due to the lack of experimental data on defect/crack nucleation and growth during stress corrosion on the microstructural level, direct validation of the model predictions is not viable. However, an indirect comparison with experimental data is performed considering the current density response as a function of immersion time. All three magnitudes of the remote tensile strain are considered ($\varepsilon^{\infty} = 0$, $\varepsilon^{\infty} = 0.05\%$, and $\varepsilon^{\infty} = 0.06\%$). The obtained results for current density are plotted in Fig. {\ref{Fig14}}. The curves in Fig. {\ref{Fig14}} are represented as the average value obtained from ten simulations for microstructures with an average grain size of 20 $\mu$m. All three cases considered exhibit a linear initial increase in current density that further increases with immersion time. The results show that the corrosion current density increases rapidly in the presence of mechanical loading. This increase in current density is directly associated with the mechanical fields that enhance the dissolution process, Eq. ({\ref{eqn32}}). The highest current density is obtained for the most stressed case ($\varepsilon^{\infty} = 0.06\%$). This trend in increasing corrosion current density in the presence of mechanical loads is consistent with experimental data conducted on the same polycrystalline material under mechanical loading \mbox{\cite{YAZDANPANAH2022, Jeong2020,Kovac2012}}. However, a direct comparison with the model predictions cannot be performed due to differences in microstructure morphology, applied overpotential, applied loading, and corrosive environment.

\subsection{Statistical analysis of grain size effect on corrosion damage} \label{sec43}
\begin{figure}[h!]
\centering
\includegraphics[width = 16cm]{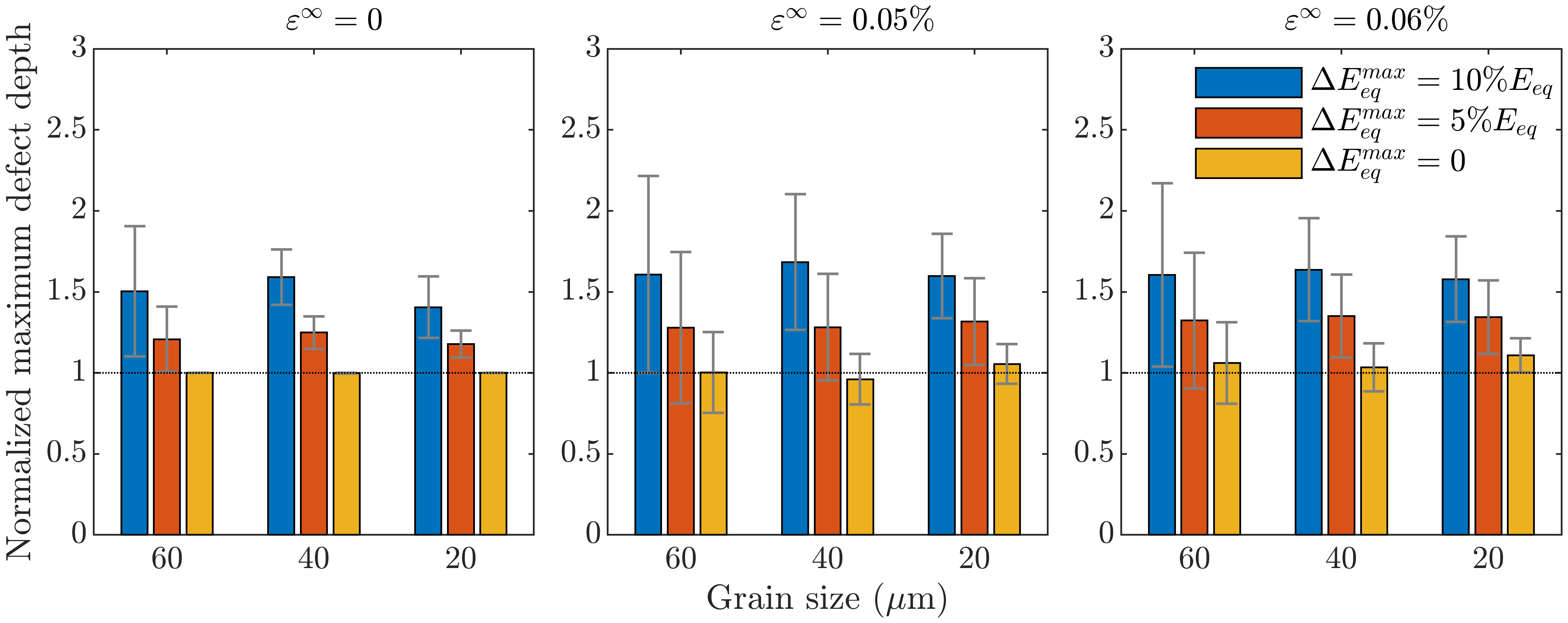}
 \captionsetup{labelfont = bf,justification = raggedright}
\caption{Normalized maximum defect depth as a function of grain size for various prescribed remote uniaxial tensile strains $\varepsilon^{\infty}$ and variations in equilibrium corrosion potential $\Delta E^{\theta}_{max}$. Each column represents an average over ten simulations and corresponding error bars show the standard deviation in each set.}
\label{Fig15}
\end{figure}

In discussing the potential of the model to assess pitting and mechanically-assisted corrosion, it is important to address the question regarding the quantitative comparison between simulations considering polycrystalline materials and homogeneous materials without the effect of microstructure. For this study, a total of two hundred seventy three phase-field simulations are performed. Ten for each of the specific conditions, three grain sizes (20 $\mu$m, 40 $\mu$m, and 60 $\mu$m), three choices of equilibrium corrosion potential variation ($\Delta E^{\theta}_{max} = 10\% E_{eq}$, $\Delta E^{\theta}_{max} = 5\% E_{eq}$, and $\Delta E^{\theta}_{max} = 0$), and three applied tensile conditions ($\varepsilon^{\infty} = 0$, $\varepsilon^{\infty} = 0.05\%$, and $\varepsilon^{\infty} = 0.06\%$) are considered in this analysis. The corresponding simulation for each applied tensile condition in the case of a homogeneous material serves as a reference study for comparison. The analysis is carried out through two metrics parameters \cite{Brewick2017}: maximum defect depth and dissolved area (change in metal area due to dissolution through corrosion). The results for both metrics parameters are normalized with respect to the homogeneous case for better comparison. Fig. \ref{Fig15} presents the normalized maximum defect depth as a function of grain size for various prescribed remote uniaxial tensile strains $\varepsilon^{\infty}$ and variations in equilibrium corrosion potential $\Delta E^{\theta}_{max}$. As depicted in Fig. \ref{Fig15}, longer maximum defects (pits and cracks) are obtained in polycrystalline materials than in homogeneous materials without the effect of microstructure. This effect is more pronounced for higher variations in equilibrium corrosion potential. Fig. \ref{Fig15} also reveals that the variation in equilibrium corrosion potential is more dominant in determining the maximum pit and crack depth than the grain size. Higher variations in corrosion potential return higher standard deviations, indicating the defect depth variation among different grain morphologies. Decreasing grain size diminishes standard deviation and returns more uniform defect depths across different grain morphologies. The anisotropic mechanical properties with a constant corrosion potential induce a certain degree of variation in defect depth. This effect tends to decrease by decreasing the grain size. Nevertheless, the results suggest that the anisotropic mechanical properties with a uniform corrosion potential on average resemble the homogeneous case. 

\begin{figure}[h!]
\centering
\includegraphics[width = 16cm]{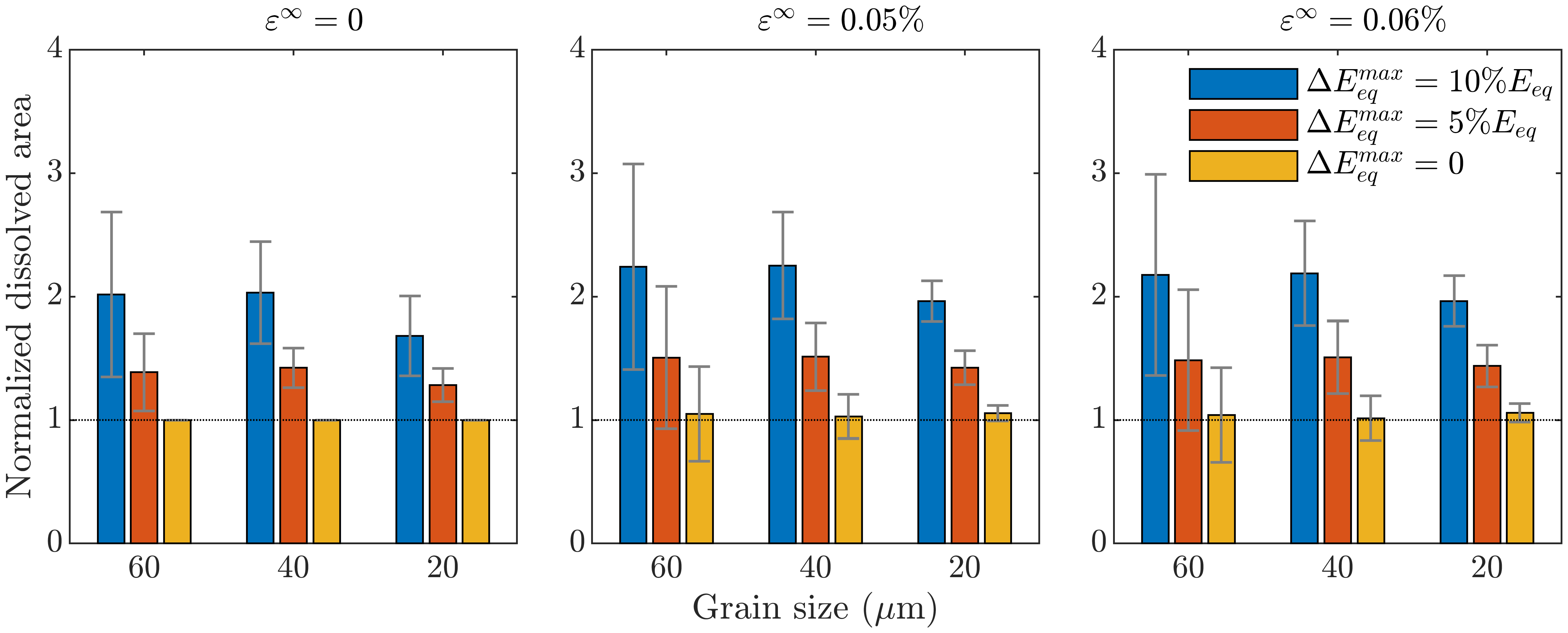}
 \captionsetup{labelfont = bf,justification = raggedright}
\caption{Normalized dissolved area as a function of grain size for various prescribed remote uniaxial tensile strains $\varepsilon^{\infty}$ and variations in equilibrium corrosion potential $\Delta E^{\theta}_{max}$. Each column represents an average over ten simulations and corresponding error bars show the standard deviation in each set.}
\label{Fig16}
\end{figure}

The impact of variations in corrosion potential on pitting ($\varepsilon^{\infty} = 0$) and stress-assisted corrosion in terms of dissolved metal area is shown in Fig. \ref{Fig16}. The results indicate that nonhomogeneous materials have significantly more severe pitting corrosion than homogeneous ones, resulting in a larger dissolved metal area. This trend is especially prominent in the case of highly nonuniform corrosion potentials and coarse microstructures. However, decreasing variations in corrosion potential and grain size tends to make the results more similar to those of a homogeneous material. The presence of mechanical loading results in a significantly larger dissolved area compared to the homogeneous case, implying that longer and wider cracks are formed, leading to more damage to the material, Fig. \ref{Fig16}. This effect shows that microstructure effects are significant even when stress-enhanced dissolution becomes the dominant failure mode. However, this trend can be mitigated by reducing the variation in corrosion potential. The anisotropic mechanical properties with a uniform corrosion potential do not result in a substantial increase in dissolved area compared to the homogeneous case.

The results obtained in Fig. \ref{Fig15} and Fig. \ref{Fig16} suggest that pitting and stress-enhanced corrosion are more pronounced when microstructural features are considered, particularly for high variations in corrosion potential and coarse microstructures. The change in strength of microstructural effects with respect to the grain size is typical of a grain and pit size competition. The effect of underlying microstructure on localized corrosion damage can be better quantified using the dissolved metal area as a metrics parameter rather than maximum defect depth. The present results demonstrate that excluding the underlying microstructure in simulating localized corrosion would underestimate the defect kinetics and corrosion damage. The current formulation may serve as a cost-effective way of providing insights into the pitting and stress corrosion cracking resistance of the material. The model can be used to predict the long-term durability of materials under various corrosive environments. The ability to predict the impact of microstructure on localized corrosion damage is appealing for practical applications in developing and tailoring more corrosion-resistant microstructures. Moreover, the model can be calibrated using only current density data, making it more attractive for practical use. Measuring current density in experiments is more feasible than tracking pit kinetics on the microscopic level. Incorporating electrochemical interfacial phenomena associated with the formation of EDL into the present model is an important step toward bridging the gap between accelerated and natural corrosion tests. 

\section{Conclusions and future work} \label{sec5}

An electro-chemo-mechanical phase-field computational framework is developed to predict pitting and stress corrosion cracking in polycrystalline materials. The model takes into account the role of microstructure by including dependencies of corrosion potential and mechanical properties on crystallographic orientation. The model considers the formation and charging dynamics of an electric double layer through a general boundary condition for the solution potential. The results indicate that the underlying microstructural features play a significant role in determining the growth and morphology of defects and the distribution of stress and plastic strain. The underlying microstructure with dependencies of corrosion potential and mechanical properties on crystallographic orientation leads to more extensive defects, faster defect kinetics, irregular pit and crack shapes, and highly nonuniform stress and plastic strain distributions when compared to a homogeneous material without the effect of microstructure. Localized corrosion damage is quantified with two metrics parameters: maximum defect depth and dissolved metal area. The results show that pitting and stress corrosion cracking is more pronounced in the presence of microstructural features, leading to longer pits and cracks, and ultimately larger dissolved metal areas. Moreover, the results reveal that crystallographic dependence and variations in corrosion potential are more dominant in determining the maximum pit and crack depth than the grain size and mismatch in mechanical properties. The present study demonstrates that excluding the underlying microstructure in simulating pitting corrosion and stress corrosion cracking would underestimate the defect kinetics and corrosion damage.

Future work should consider the role of grain boundaries, surface roughness of metal surfaces, and dependence of interfacial energy on grain orientation in pitting corrosion and stress corrosion cracking. Crystal plasticity can be readily incorporated into the present framework.

\section*{Acknowledgments} \label{sec7}

M.M., M.R.W. and E.M.-P. acknowledge support from the EPSRC Centre for Doctoral Training in Nuclear Energy Futures [Grant EP/5023844/1]. S.K. and E.M.-P. acknowledge financial support from UKRI’s Future Leaders Fellowship program [Grant MR/V024124/1]. 

\section*{Data Availability} \label{sec8}

The code developed will be available at \url{www.imperial.ac.uk/mechanics-materials/codes} after article acceptance.

\appendix
\setcounter{figure}{0}
\renewcommand{\thefigure}{A.\arabic{figure}}

\section*{Appendix A. Mesh sensitivity analysis and grain morphology distributions} \label{appendixA}
\subsection*{A.1. Mesh sensitivity analysis} \label{appendixA1}

The mesh sensitivity study is performed for the pencil electrode test described in Section {\ref{sec3}}. The maximum element size in the interface propagation region is defined relative to the interface thickness $\ell$. A convergence analysis is performed in which the maximum element size is varied. Three cases are considered for the maximum element size: $\ell/2$, $\ell/4$, and $\ell/6$. Triangular finite elements with second-order Lagrangian interpolation functions are used in the simulations. The maximum time increment is constrained to $\Delta t$ = 0.2 s and a relative solver tolerance of $10^{-4}$ is used. Refer to Section {\ref{sec25}} for more details regarding numerical implementation.

\begin{figure}[h!]
\centering
\includegraphics[width = 16cm]{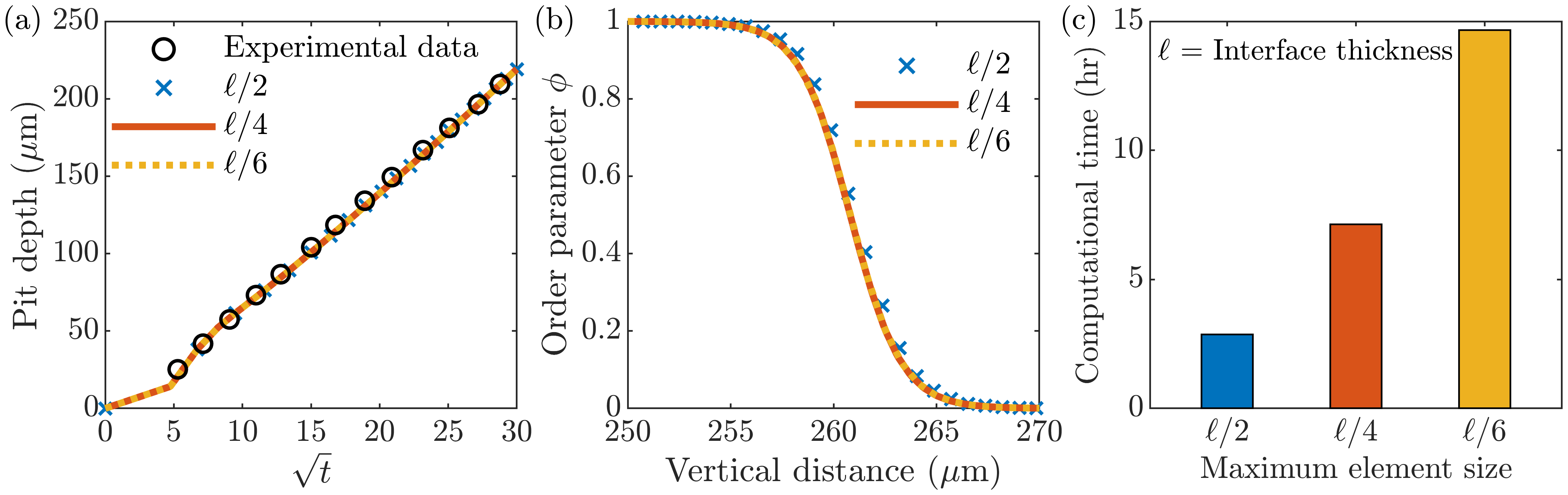}
 \captionsetup{labelfont = bf,justification = raggedright}
\caption{The comparison of (a) pitting kinetics (a), (b) interface profile, and (c) computational time as a function of maximum element size across the diffuse interface thickness. The maximum element size used in all simulations in this investigation is set to $\ell/4$. The interface profile in (b) is given along the symmetry axis (Fig. \ref{Fig4}) at the final computational time.}
\label{A1}
\end{figure}

Under those selected element types and solver settings, all three cases considered for the maximum element size returned the same pitting kinetics, Fig. {\ref{A1}}(a). Moreover, the interface profile at the final computational time for the $\ell/4$ case coincides with the $\ell/6$ case, while a slight discrepancy exists compared to the $\ell/2$ case, Fig. {\ref{A1}}(b). Therefore, increasing the number of elements across the interface thickness to more than four does not improve the results while the computational time significantly increases, Fig. {\ref{A1}}(c). The maximum element size used in all simulations in Section {\ref{sec3}} and Section {\ref{sec4}} is selected based on this convergence analysis and set to $\ell/4$.

\subsection*{A.2. Grain morphology distributions} \label{appendixA2}

\begin{figure}[h!]
\centering
\includegraphics[width = 15cm]{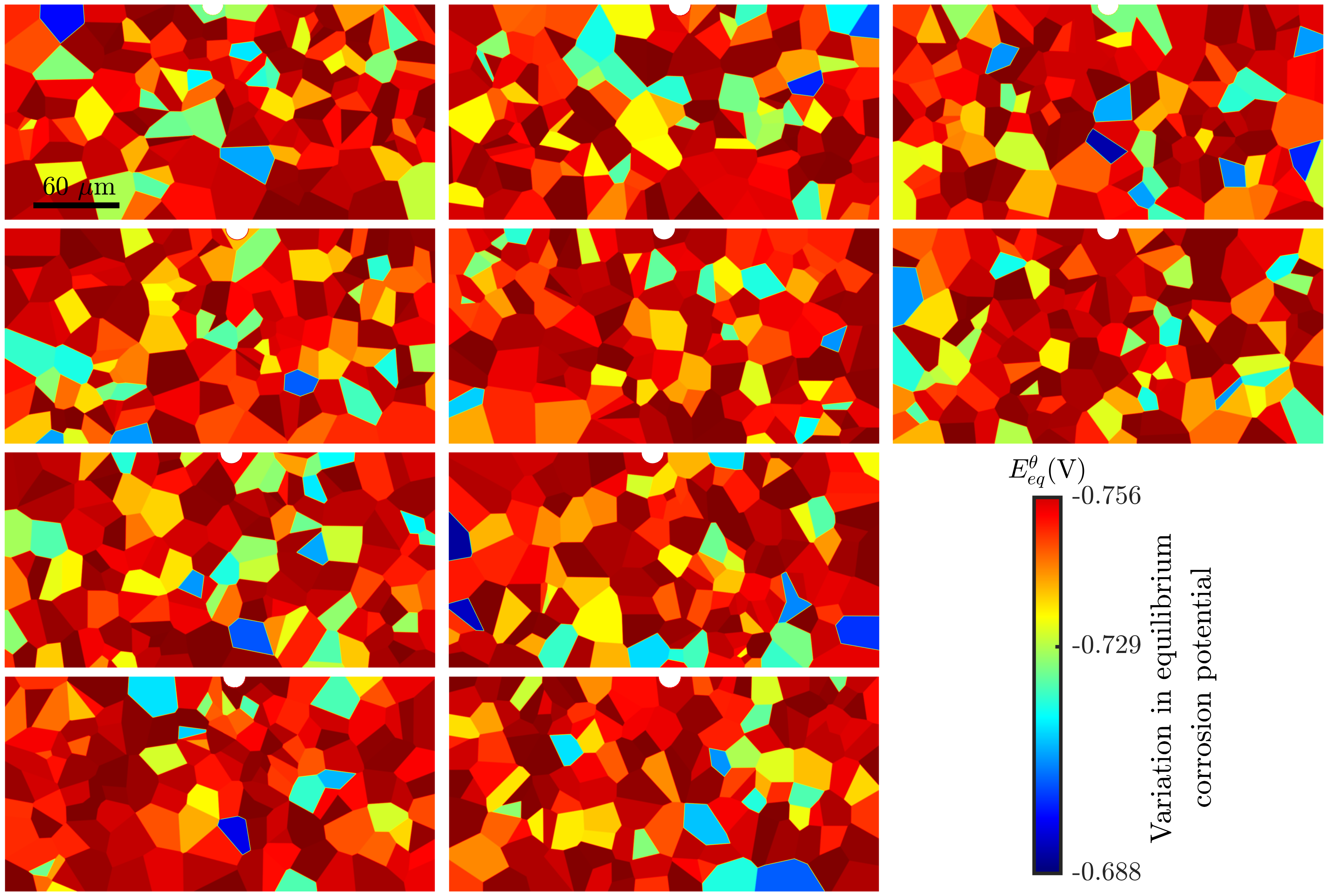}
 \captionsetup{labelfont = bf,justification = raggedright}
\caption{Grain morphology distributions for the average grain size of 20 $\mu$m. The color bar represents variation in equilibrium corrosion potential with $\Delta E^{\theta}$ = 10\%$E_{eq}$ (vs. SCE) and applies to all figures. $E_{eq} = -0.729$ V (vs. SCE) is the macroscopic equilibrium corrosion potential. The width and height of the polycrystalline material are 300 $\mu$m and 150 $\mu$m, Fig. \ref{Fig8}. The surrounding corrosive environment is not shown in these figures.}
\label{A2}
\end{figure}

\begin{figure}[H]
\centering
\includegraphics[width = 15cm]{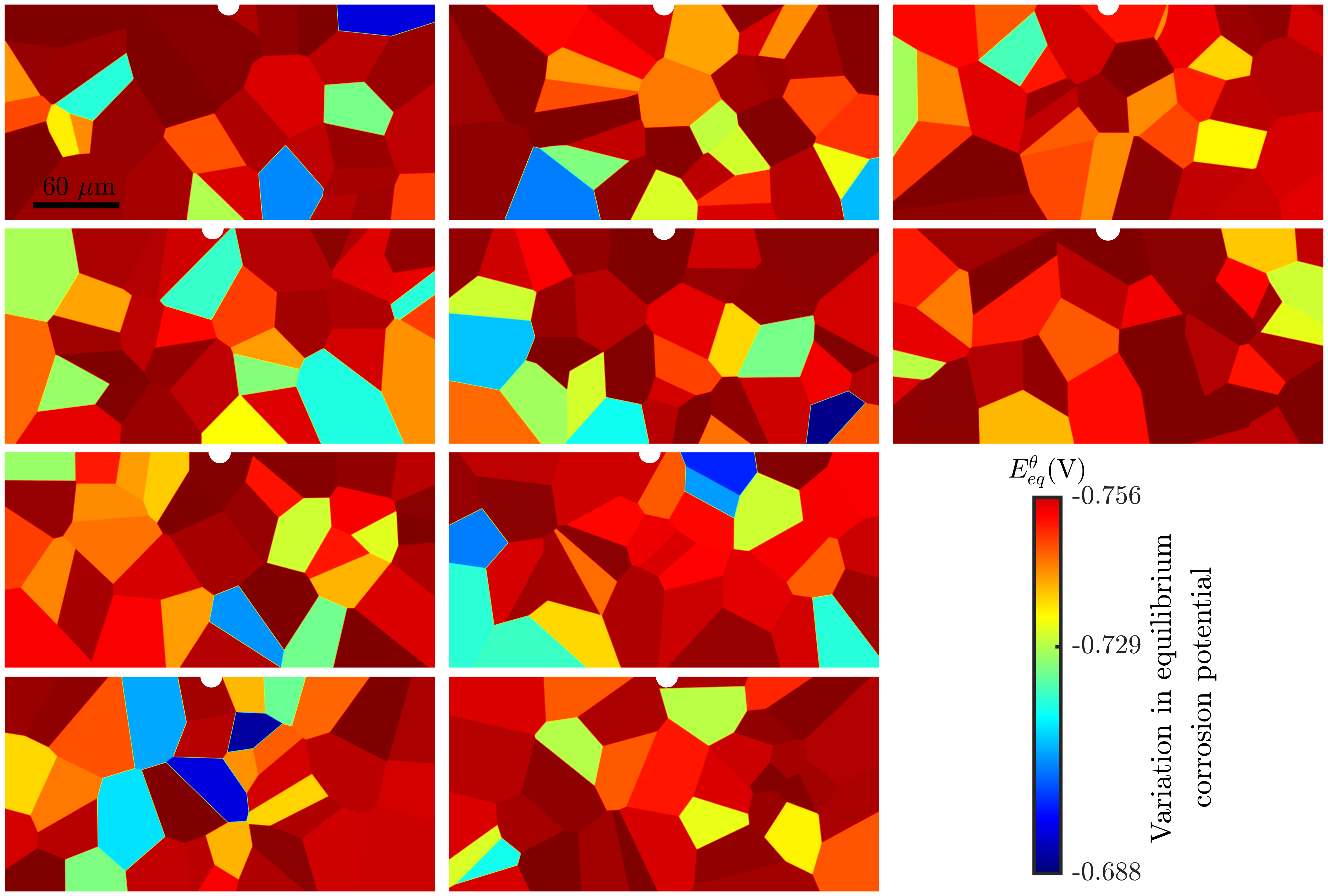}
 \captionsetup{labelfont = bf,justification = raggedright}
\caption{Grain morphology distributions for the average grain size of 40 $\mu$m. The color bar represents variation in equilibrium corrosion potential with $\Delta E^{\theta}$ = 10\%$E_{eq}$ (vs. SCE) and applies to all figures. $E_{eq} = -0.729$ V (vs. SCE) is the macroscopic equilibrium corrosion potential. The width and height of the polycrystalline material are 300 $\mu$m and 150 $\mu$m, Fig. \ref{Fig8}. The surrounding corrosive environment is not shown in these figures.}
\label{A3}
\end{figure}

\begin{figure}[H]
\centering
\includegraphics[width = 15cm]{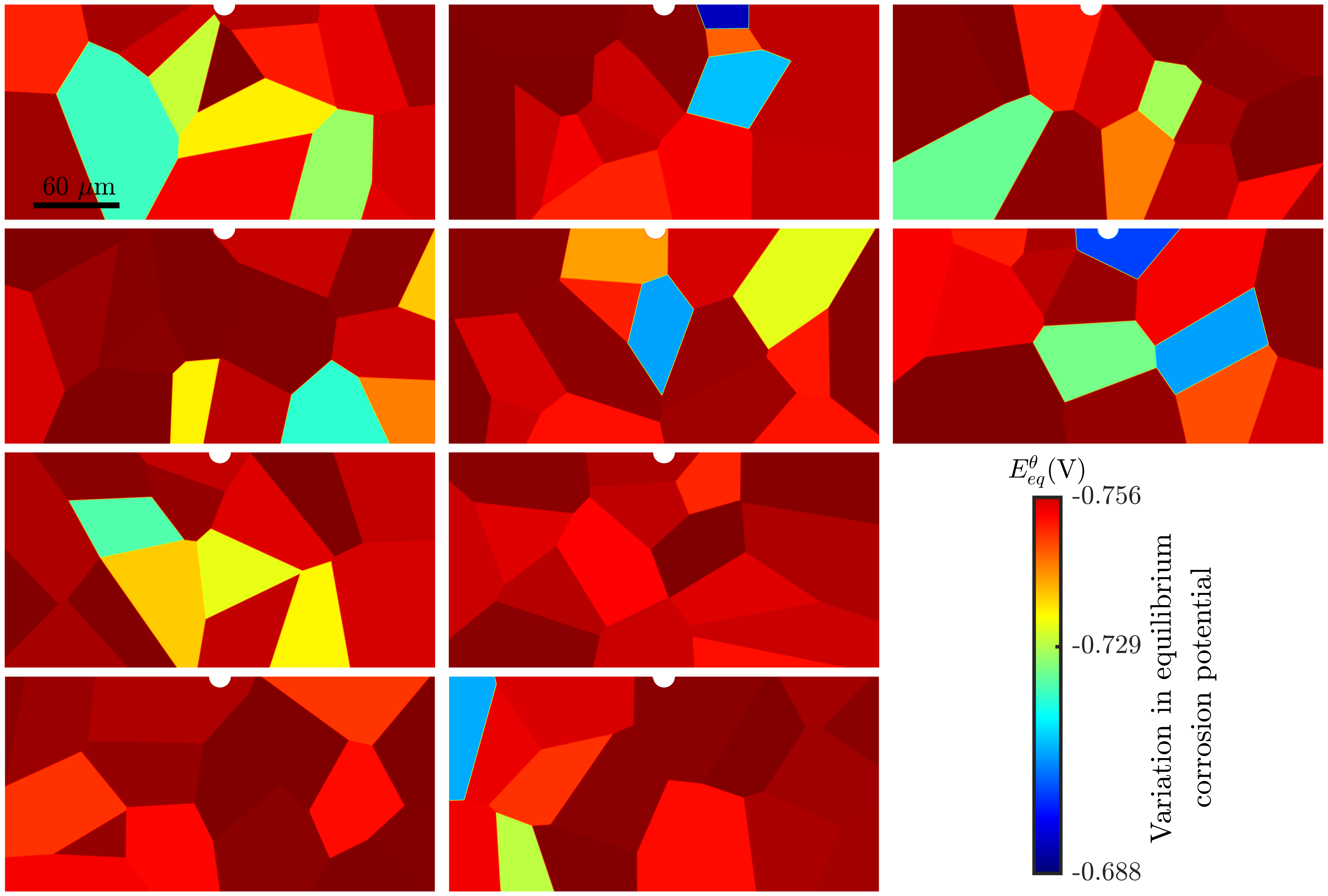}
 \captionsetup{labelfont = bf,justification = raggedright}
\caption{Grain morphology distributions for the average grain size of 60 $\mu$m. The color bar represents variation in equilibrium corrosion potential with $\Delta E^{\theta}$ = 10\%$E_{eq}$ (vs. SCE) and applies to all figures. $E_{eq} = -0.729$ V (vs. SCE) is the macroscopic equilibrium corrosion potential. The width and height of the polycrystalline material are 300 $\mu$m and 150 $\mu$m, Fig. \ref{Fig8}. The surrounding corrosive environment is not shown in these figures.}
\label{A4}
\end{figure}

Ten different synthetically generated microstructures for each grain size considered in this work (20 $\mu$m, 40 $\mu$m, and 60 $\mu$m) are given in Figs. ({\ref{A2}}), ({\ref{A3}}), and ({\ref{A4}}). Three Euler angles $\varphi_1$, $\varphi_2$, and $\varphi_3$, which are randomly assigned to each grain, are used to determine the dependence of equilibrium corrosion potential on crystallographic orientation in Eq. ({\ref{eqn31}}). The dependence of equilibrium corrosion potential on grain orientation is given in Fig. {\ref{Fig3}}.

\begin{singlespace}

\end{singlespace}
\end{document}